\documentclass[prper,aps,twocolumn,groupedaddress,floats,showpacs,final,superscriptaddress]{revtex4-1}
\usepackage{leading}
\usepackage[latin3]{inputenc}
\usepackage[makeroom]{cancel}
\usepackage{graphicx}
\usepackage{amsmath}
\usepackage{amsfonts}
\usepackage{amssymb}
\usepackage{color}
\usepackage[left=2cm,right=2cm,top=2cm,bottom=2cm]{geometry}
\usepackage{graphicx} 
\usepackage{dcolumn}
\usepackage{bm}
\usepackage{simplewick}
\usepackage{array}
\usepackage{appendix}

\RequirePackage[
   hyperindex,colorlinks,bookmarksnumbered,
   plainpages=true,pdfstartview=FitH]{hyperref}
\hypersetup{linkcolor=blue,urlcolor=blue,citecolor=blue}
\usepackage{hyperref}
\usepackage{mathrsfs}
\definecolor{purple}{rgb}{0.5,0,0.6}

\usepackage{ulem}
\renewcommand{\emph}[1]{\textit{#1}}
\definecolor{darkblue}{rgb}{0,0,0.5}
\definecolor{darkgreen}{rgb}{0,0.5,0}
\definecolor{darkred}{rgb}{.7,0,0}
\definecolor{purple}{rgb}{0.5,0,0.6}
\definecolor{orange}{rgb}{1,0.5,0}
\definecolor{grey}{rgb}{.6,.6,.6}
\definecolor{lightpink}{rgb}{1,0.7,0.75}
\definecolor{pink}{rgb}{1,0.4,0.58}
\definecolor{deeppink}{rgb}{1,0.08,0.58}

\newcommand{\DK}[1]{{\color{black}{#1}}} 

\renewcommand{\emph}[1]{\textit{#1}}

\newcommand{\cp}{\cos2\delta_{\rm P}}
\newcommand{\cs}{\sin2\delta_{\rm P}}
\newcommand{\rss}{\mathbb{S}}

\newcommand{\Tk}{T_{\rm K}}

\newcommand{\ed}{\varepsilon_{\rm d}}
\newcommand{\ma}{\mathscr{A}}
\newcommand{\mb}{\mathscr{B}}
\newcommand{\oa}{\overline{\alpha}}
\newcommand{\ot}{\overline{T}}
\newcommand{\mt}{\mathcal{T}}

\begin{document}
\title{Transmission phase evolution in fully screened and overscreened Kondo impurities}

\author{D. B. Karki}
\affiliation{Division of Quantum State of Matter, Beijing Academy of Quantum Information Sciences, Beijing 100193, China}

\begin{abstract}
\DK{We study} the coherent properties of fully screened and overscreened Kondo effects based on the Nozieres local Fermi-liquid theory and Affleck-Ludwig boundary conformal-field-theory approach, respectively. \DK{Coherent transports through an SU($N)$ generalization of
fully screened and the multi-$\mathcal{K}$-channel overscreened Kondo impurities at and beyond the particle-hole (PH) symmetric point are thoroughly investigated}. We report distinctive Fermi-liquid coefficients characterizing the finite temperature correction to the transmission phase shift and normalized visibility in fully screened Kondo regime, which can be measured with the existing experimental setups. Our work equally uncovers the significance of temperature correction to the transmission phase shift and visibility in non Fermi-liquid regime associated with the \DK{PH} asymmetric overscreened Kondo effects with an arbitrary number of conduction channels $\mathcal{K}$. We propose viable roots of verifying our predictions in connection to the recent experiments, in particular with the experiments studying highly symmetric forms of fully screened Kondo effects and two-channel overscreened Kondo effects realized in quantum dot nano-structures.
\end{abstract}

\date{\today}


\maketitle
\section{Introduction}
The Kondo effect~\cite{Kondo} involves the coupling of $\mathcal{K}$ degenerate channels of conduction electrons to the spin $\vec{\mathcal{S}}$ of the quantum impurity~\cite{Anderson, haman1, Anderson_Yuval_Hamann, Wilson,jvm1, Andrei_RevModPhys_1983,AFFLECK_NPB_1990,Affleck_Lud_PRB(48)_1993}. The size of effective spin $\mathcal{S}$ and number of conduction channels $\mathcal{K}$ define the different regimes of the Kondo problem. On of the widely studied class of Kondo paradigm is the fully screened Kondo effects with $\mathcal{K}=2\mathcal{S}$~\cite{Nozieres}. While the condition $\mathcal{K}<2\mathcal{S}$ results an underscreened Kondo effect, the highly non-trivial version of Kondo effects are the overscreened one satisfying $\mathcal{K}>2\mathcal{S}$~\cite{Nozieres_Blandin_JPhys_1980, Andrei_RevModPhys_1983, Sacramento_CM(48)_1991, Cox_Adv_Phys(47)_1998}. A local Fermi-liquid (FL) theory pioneered by Nozieres successfully explains the low energy transport features of the Kondo effect in fully screened and underscreened regime~\cite{Nozieres, Goldhaber_nat(391)_1998}. Nevertheless, the low lying excitations in overscreened Kondo effects are not \DK{electron-like} quasiparticles as in FL. Consequently, the overscreened case possess strikingly different behavior with non-FL (NFL) ground state~\cite{Affleck_Lud_PRB(48)_1993,Nozieres, Nozieres_Blandin_JPhys_1980, yuval_david,delft2, Potok_NAT(446)_2007}.

One of the elegant result form the seminal work of Nozieres~\cite{Nozieres} is the description of fully screened Kondo effects in terms of the scattering phase shifts between incoming and outgoing states. As an examples, the quantum impurity with an effective spin-1/2 interacting with the single conduction channel $\mathcal{K}{=}1$ results a zero temperature $\pi/2$ phase shift between the incoming and outgoing states. This is a very nature of elastic scattering in FL~\cite{Hewson}. Nonetheless, in overscreened $\mathcal{K}$-channel Kondo ($\mathcal{K}$CK) effects the situation becomes completely different, namely the $\mathcal{K}$CK impurity mediates the single particle to many-body state scattering processes~\cite{Affleck_Lud_PRB(48)_1993, carmi1}.

In the recent years, the different experimental works devoted to the transport measurements in $\mathcal{K}$CK has further revive the field of Kondo paradigm. In particular, the experimental success of  transport measurement in 1CK with two spin flavors~\cite{Goldhaber_nat(391)_1998, kiselev, paulo, heiner, new} has recently been extended to those with $N$-spin flavors~\cite{jh, sasa0, sasa, if2, ll1,keller} supplemented by various theoretical \DK{works}~\cite{hur,su4_21, su4_23,Lim, su4_11, dee1, last1, den}. Likewise, the celebrated experimental observation of spin 2CK physics~\cite{Potok_NAT(446)_2007} has eventually leads to the success of recent experimental observation of charge 2CK and 3CK effects~\cite{ Pierre_NAT(526)_2015, iff}. In addition, different experiments on the Aharonov-Bohm (AB) interferometry provide an elegant way of measuring coherent properties of 1CK effects~\cite{moti_hiblum,moti_1, moti_2008, delft_andrea}, namely the transmission phase shift (TPS) and the normalized visibility measurement.

The recent work of Ref.~\cite{carmi1} has provided a theoretical framework for the study of TPS for two spin-flavored Kondo impurities in both the FL and NFL regime. The assumed $SU(2)$ spin symmetric Kondo impurities are, by construction, the particle-hole (PH) symmetric. The very essence of Kondo effect in PH symmetric situation is the trivial $\pi/2$ phase shift without any temperature corrections even beyond the Kondo temperature $\Tk$~\cite{Affleck_Lud_PRB(48)_1993}. On one hand, in many practical situations the PH symmetry might be broken either in the impurity or in the conducting reservoirs. On the other hand, certain class of Kondo impurities might not even posses the PH symmetric ground state. The example includes, the seminal work~\cite{jh,keller} on 1CK $SU(4)$ Kondo effects which have both PH symmetric and beyond PH symmetric ground states depending on the electron occupancy. In order to study the complete evolution of TPS of Kondo paradigmatic phenomena one, therefore, needs a theoretical description which can consistently account for the effects of PH symmetry breaking (lifting the PH symmetry). 

In this work we develop a full-fledged theoretical framework to study the coherent properties of fully screened and overscreened Kondo effects. The fully screened Kondo effects under consideration are with single channel $\mathcal{K}{=}1$ and $N$-spin flavors. These impurities posses an exact $SU(N)$ spin symmetry on the top of $U(1)$ charge symmetry~\cite{Affleck_Lud_PRB(48)_1993} and for the sake of simplicity we refer them as $SU(N)$ Kondo impurities throughout the text. The overscreened Kondo effects we consider in this work are those with an effective spin-1/2 and arbitrary number of conduction channels $\mathcal{K}{\geq}2$ with the symmetry $SU(2){\times}SU(\mathcal{K}){\times}U(1)$~\cite{Nozieres_Blandin_JPhys_1980,Affleck_Lud_PRB(48)_1993}. In addition, we describe the 1CK impurities in $SU(N)$ symmetric regime in terms of Nozieres local FL approach~\cite{Nozieres}, while the corresponding description for the overscreened $\mathcal{K}$CK effects would be based on the boundary conformal-field-theory (BCFT) approach for the Kondo problem originally formulated by Affleck and Ludwig (AL)~\cite{AFFLECK_NPB_1990,Affleck_Lud_PRB(48)_1993}. We emphasized that the ideas presented in this work are not limited to the study of coherent properties, they could be equally useful for other measure of quantum transport such as the thermoelectric transport.

In the following we summarize our main findings. Using a local Fermi liquid theory and the physics of the AB-effect, we investigate the finite temperature TPS $\theta_{\rm tp}$ and normalized visibility $\eta$ of a quantum impurity fine tuned to the fully screened $SU(N)$ Kondo regime. Our result for the TPS with electron occupancy \DK{$m=1, 2, \cdots N-1$} reads~\footnote{The presented Fermi-liquid description is limited to the single impurity Anderson model with symmetric bath density of states.} \DK{(see next section for the details of experimental setup and explicit definition of TPS)}
\begin{equation}\label{llm1}
\left.\theta_{\rm tp}\right|_{m, N}=\frac{m\pi}{N}+\mathscr{C}_{{\rm T},N}(m) \left(\frac{\pi T}{\Tk}\right)^2,
\end{equation}
where $\Tk$ is the Kondo temperature and $\mathscr{C}_{{\rm T},N}(m)$ stands for the FL coefficient characterizing the finite temperature correction
\begin{equation}
\mathscr{C}_{{\rm T},N}(m)=\frac{1}{3}\frac{N+1}{N-1}\cot \left(\frac{\pi  m}{N}\right).
\end{equation}
This result indicates that the TPS permits the finite temperature $T\ll\Tk$ correction beyond PH symmetric limit. We propose to use the coefficient characterizing the finite temperature correction $\mathscr{C}_{{\rm T},N}(m)$ as a distinctive FL coefficient (identity) that can me measured with the existing experimental setup~\cite{delft_andrea}. 

The another central result of this work is the investigation of normalized visibility $\eta$ which represents the amount of coherent transport in $SU(N)$ Kondo regime. Nozieres FL based calculations results in the finite temperature correction to $\eta$ as a function of $m$ and $N$ as given by
\begin{equation}\label{dfr}
\left.\eta\right|_{m, N}=1-\frac{1}{3\sin^2\left(\frac{m\pi}{N}\right)}\;\frac{N+1}{N-1}\;\left(\frac{\pi T}{\Tk}\right)^2.
\end{equation}
The equation~\eqref{dfr} expresses the general form of finite temperature reduction of coherent transport in $SU(N)$ Kondo regime with electron occupancy $m$. At zero temperature the incoming electron only scatters into a single outgoing electron resulting in the completely coherent transport~\cite{affjs}. At finite temperature situation in FL, however, the scattering phase shift acquires the energy dependence and also the inelastic processes become active resulting thereby reducing the amount of coherent transport as expressed in Eq.~\eqref{dfr}.

From equation~\eqref{llm1} it is seen that the special case of $N=2 m$ (PH symmetric cases) $SU(N)$ Kondo variants have the trivial $\pi/2$ phase shifts between the incoming and the outgoing electron states. The $\pi/2$ phase shift is the consequences of the PH symmetry which is only true if the system is fine tuned (e.g. by gate voltage) to the middle of Coulomb blockade valley. Therefore, of the particular experimental interests, we explore the gate voltage dependence of TPS in $SU(2)$ Kondo regime. Our result for TPS reads
\begin{equation}
\left.\theta_{\rm tp} \right|_{SU(2)}=\delta_0+ \left[\frac{2}{9}\;\overline{\alpha}_2+\overline{\alpha}^2_1\cot\delta_0\right]\left(\frac{\pi T}{\mathcal{F}}\right)^2,
\end{equation}
where $\delta_0$ is the zero energy phase shift and $\overline{\alpha}_{1, 2}$ are FL parameters. These parameters $\delta_0$ and $\overline{\alpha}_{1, 2}$ depend on the original parameters of the Anderson model, namely the dot energy level $\varepsilon_{\rm d}$
level hybridization $\mathcal{F}$ and the Coulomb energy $U$. At the middle of Coulomb-blockade valley where the PH symmetry is restored $\delta_0{=}\pi/2$ and $\overline{\alpha}_2{=}0$ and hence reproducing the usual $\pi/2$ phase shift as one of the hallmark of $SU(2)$ Kondo physics.

Soon after completing the FL description of TPS evolution in 1CK effects, we develop AL CFT based description for the TPS evolution in $\mathcal{K}$CK effects. For fixed number of conduction channel $\mathcal{K}$, our main result for TPS in the low temperature regime $T\ll\Tk$ of the overscreened Kondo regime depends on two parameters, the Kondo temperature $\Tk$ and the potential scattering phase shift $\delta_{\rm P}$
\begin{equation}\label{manb1}
\theta_{\rm tp}{=}{\tan^{-1}}\!\Bigg[\!\frac{1{-}\cp \frac{\cos \left(\pi\Delta\right)}{
\cos \left(\pi\Delta/2\right)}{-}\cp\mathscr{D}(\Delta)\!\left(\frac{\pi T}{\Tk}\right)^{\Delta}}{\cs\frac{\cos \left(\pi\Delta\right)}{
\cos \left(\pi\Delta/2\right)}+\cs\mathscr{D}(\Delta)\!\left(\frac{\pi T}{\Tk}\right)^{\Delta}}\!\!\Bigg],
\end{equation}
with $\Delta\equiv 2/(2+\mathcal{K})$ and $\mathscr{D}(\Delta)$ stands for some constant for given $\mathcal{K}$. At PH symmetric point, owing to have $\delta_{\rm P}=0$, the denominator of Eq.~\eqref{manb1} incidentally vanishes resulting in the $\pi/2$ phase shift as in conventional Kondo problems. However, the complete evolution of TPS in NFL regime given by Eq.~\eqref{manb1} is much more interesting than that of corresponding FL regime. In addition, the temperature correction to the TPS is controlled by the PH asymmetry parameter $\delta_{\rm P}$ and the fractional temperature exponent $\Delta$ unlike the $T^2$ correction in conventional FL regime.

Before closing this introductory section we discuss briefly on the organization of this paper. In Sec.~\ref{sec1}, we discuss about the coherent properties of the system that can be studied using the AB-interferometry. In addition, we introduce the fundamental mathematical identity relating the spectral function and the coherent properties. The Sec.~\ref{sec2} is devoted to construction of low-energy theoretical framework for the TPS evolution of Kondo impurities in FL regime based on the Nozieres seminal work~\cite{Nozieres}. In this section we discuss the general behavior of finite temperature TPS and the normalized visibility in PH symmetric and beyond PH symmetric cases separately. The detail on the mathematical formulation and discussion of the coherent properties in overscreened Kondo regime is presented in Sec.~\ref{sec3}. Finally, we conclude our work in Section~\ref{sec4}.

\section{AB-effect and coherent properties}\label{sec1}
The TPS is one of the characteristic feature of Kondo correlation which, however, can not be measured directly in conventional Kondo setup. The AB ring containing the quantum impurity in one of the arm has been suggested~\cite{oreg_delft} as an effective tool of obtaining the TPS information. The TPS has been shown to be easily extracted from the AB oscillations of the conductance as a function of the magnetic flux through the AB ring~\cite{delft_andrea, oreg_delft}.

In a typical AB interferometer two different electronic paths are connected between source S and drain D. The quantum impurity under study is then placed in one of the path (let us say lower) and the remaining one (upper) is the reference path. The transport measurement of one path can be done independently of other by switching off the later~\cite{carmi1}. When an electron is send from source it suffers a path difference of $\Phi e/\hbar c+\theta_{\rm tp}$ while reaching to the drain due to AB effect. Thus the interference between two paths depends on the magnetic flux $\Phi$ enclosed within the AB ring and the TPS $\theta_{\rm tp}$ which can be regarded as the relative phase between two paths. The source-drain conductance $G_{\rm SD}$ reads~\cite{carmi1}
\begin{equation}
G_{\rm SD}=G_{\rm d}+G_{\rm rf}+2\sqrt{\eta}\sqrt{G_{\rm d}G_{\rm rf}}\cos\left(\Phi e/\hbar c+\theta_{\rm tp}\right),
\end{equation}
where $G_{\rm d}$ is the conductance of the corresponding path with the QD (impurity of interest) and $G_{\rm rf}$ is the conductance of reference path. The factor $\eta$ is a constant characterizing the amount of coherent transport as compared to the total transport. 

Following the setup proposed in Ref.~\citep{carmi1}, we now briefly describe the electron transport in open AB ring. The setup contains two external leads acting as the source and drain and two internal paths. While the upper path is the reference path, the lower path further contains two internal leads (the left and right leads) connecting the impurity of interest. The transmitted electron from the source to the left leads then propagate toward the impurity. These electrons get scattered off the impurity and propagate along the right lead which finally get transmitted out into the drain. Given that the source and drain are very weakly coupled to the internal leads~\citep{carmi1}, the strong electron interaction on the top of resonance scattering can be realized in the left lead-impurity-right lead system.

The low energy transport properties, such as the conductance, of the system mainly governed by the corresponding spectral function or the T-matrix. We connect the spectral function $\left[-\pi\nu \mathcal{T}^{\rm tot}_r (\varepsilon, T)\right]$ and the TPS by defining the thermally and spin averaged function
\begin{equation}\label{sita1}
\mathcal{J}(T)=\frac{1}{4T}\sum_{r}\int^{\infty}_{-\infty}\;\frac{d\varepsilon}{2\pi}\times \frac{-\pi\nu \mathcal{T}^{\rm tot}_r (\varepsilon, T)}{\cosh^2 \left(\frac{\varepsilon}{2T}\right)},
\end{equation}
where we write the derivative $\left(-\partial f(\varepsilon, T)/\partial\varepsilon\right)$
of equilibrium Fermi distribution function $f(\varepsilon, T)$ in terms of trigonometric function. In addition, the symbol $r$ stands for the spin index with $\nu$ being density of states per spin per channel. While the imaginary part of Eq.~\eqref{sita1} provides the conductance $G_{\rm d}$ of the quantum impurity, the TPS $\theta_{\rm tp}$ reads from the argument of Eq.~\eqref{sita1}~\cite{carmi1}
\begin{equation}\label{sita2}
G_{\rm d}(T)={\rm Im}\mathcal{J}(T),\;\;\theta_{\rm tp}(T)=\arctan\Bigg[\frac{{\rm Im}\mathcal{J}(T)}{{\rm Re}\mathcal{J}(T)}\Bigg].
\end{equation}
Here ${\rm Im}\mathcal{J}$ and ${\rm Re}\mathcal{J}$ stand respectively for the imaginary and real part of $\mathcal{J}$. We make the use of Eq.~\eqref{sita2} to unveil the beyond PH symmetric evolution of TPS in different variants of $\mathcal{K}$CK effects \DK{while the TPS for corresponding PH symmetric system always locked to $\pi/2$ since ${\rm Re}\mathcal{J}(T)=0$ (See next section for details).}

In addition, the another quantity of paramount experimental interests is the normalized visibility $\eta$ which accounts for the proportion of coherent transport over the total transport for given experimental setup. The Eq.~\eqref{sita1} paves the direct way of writing an expression of $\eta$~\cite{carmi1}
\begin{equation}\label{doka1}
\eta(T)={\rm Im}\mathcal{J}(T)+\Bigg[\frac{{\rm Re}\mathcal{J}(T)}{\sqrt{{\rm Im}\mathcal{J}(T)}}\Bigg]^2.
\end{equation}
For the Kondo correlated nano devices in PH symmetric limit, the transmission coefficient is fully imaginary resulting in the vanishing of the coefficient ${\rm Re}\mathcal{J}(T)$. In this situation the normalized visibility is equivalent to the differential conductance. Beyond PH symmetric point, the spectral function, however, becomes asymmetric with finite ${\rm Re}\mathcal{J}(T){\neq}0$. We explore in detail the finite temperature behavior of $\eta$ expressed in Eq.~\eqref{doka1} for both the fully screened and overscreened Kondo effects at the nano scale.

\section{Fully screened Kondo effects}\label{sec2}
The Kondo paradigmatic phenomena satisfying the condition $\mathcal{K}=2\mathcal{S}$ belongs to the class of fully screened Kondo effects~\cite{Nozieres_Blandin_JPhys_1980}. The prototypical example of fuly screened Kondo effects includes the spin-1/2 impurity interacting with the single conduction channel $\mathcal{K}=1$. In addition, the class of $\mathcal{K}=2\mathcal{S}$ can straightforwardly be extended to the multi-channel scenario with higher spin $\mathcal{S}>1/2$~\cite{HWDK_PRB_(89)_2014, dee2, dee3}. Since the multi-channel version of fully screened Kondo effects have fundamentally the similar properties as per corresponding single channel, in this work we restrict ourselves to the latter case. In addition, the single channel Kondo paradigm also offers the generalization of the corresponding ground states from the conventional $SU(2)$ symmetric to the ground state with exotic $SU(N\geq 2)$ symmetry. The orbital degeneracy of the quantum impurity often combines with the true spin symmetry to form the Kondo
effect described by higher symmetry group $SU(N)$~\cite{jh, sasa0, sasa, if2, ll1,keller}. In these fully screened Kondo effects with higher symmetry group, the electron occupancy factor $m$ takes all possible values starting from 1 to $N-1$ thereby allowing the rich interplay of $N$ and $m$~\footnote{We note that the presented work describes the case of filling factor $m/N<1$ with $m=1, 2,\cdots N-1$.}. 

As is clear from the previous section that our goal in this section is to investigate the TPS in $SU(N)$ symmetric regime of Kondo effects, we first need to find the real and imaginary parts of the T-matrix. The low energy form of the T-matrix mainly comes from the Hamiltonian description which is the subject of following subsection.

\subsection{Low energy Hamiltonian for $SU(N)$ Kondo impurity and the T-matrix}
To get a close expression of T-matrix characterizing the low energy physics of $SU(N)$ Kondo effects, first we consider a quantum impurity tunnel coupled to two conducting reservoirs. By construction, we assume that the impurity possess $N$-fold degeneracy by combining the spin and other degrees of freedom, such as the orbital degeneracy. The important point to mention is the fact that the $SU(N)$ physics can be observed only if the electrons in both reservoirs also possesses the $SU(N)$ symmetry so that the rotation of the reservoir's electrons is described by the $SU(N)$ transformation. The generic model for the description of such $SU(N)$ impurity is the Anderson model with $N$ flavors~\cite{and1}. However, at low energy regime the Anderson model can be directly mapped to the local FL Hamiltonian pioneered by Nozieres~\cite{Nozieres}.

The first procedure towards the local FL description of $SU(N)$ Kondo effects involves the
rotation of the electron operators in the left and right leads $c_{\rm L, R}$ by the Glazman-Raikh rotation~\cite{GP_Review_2005} (we consider the symmetric lead-impurity coupling)
\begin{equation}\label{sane5}
 \left(%
\begin{array}{c}
  b\\
  a\\
\end{array}%
\right)= \mathbb{U}\left(%
\begin{array}{c}
  c_{{\rm L}} \\
  c_{{\rm R}} \\
\end{array}%
\right),\;\;\mathbb{U}\equiv\frac{1}{\sqrt{2}}\left(%
\begin{array}{cc}
 1 & \phantom{-}1 \\
  1 & -1 \\
\end{array}%
\right).
\end{equation}
For 1CK effects, the transformation Eq.~\eqref{sane5} effectively decouples the operators $a$ from the impurity degrees of freedom~\cite{dee1, last1} while the operator $b$ remains interacting with impurity. In other word, while both of the operators $a$ and $b$ determines the non-interacting Hamiltonian, the scattering and interaction involves merely the operator $b$. Afterwards, the scattering (the elastic effects) and the interactions (the inelastic effects) in FL is accounted for by phenomenological parameters $\alpha_{1, 2}$ and $\phi_{1, 2}$ respectively to write the low energy FL Hamiltonian~\cite{Nozieres, mora1, mora2}
\begin{eqnarray}
\mathscr{H}_{0}&=&\nu\sum_{\rm r}\int_{\varepsilon}  \varepsilon \left[ a^{\dagger}_{\varepsilon \rm r} a_{\varepsilon \rm r}+b^{\dagger}_{\varepsilon \rm r} b_{\varepsilon \rm r}\right]\label{HamFL},\\
\mathscr{H}_{\rm el}&=&-\sum_{\rm r}\int_{\varepsilon_{1-2}}\left[\frac{\alpha_1}{2\pi}(\varepsilon_1{+}\varepsilon_2){+}\frac{\alpha_2}{4\pi}(\varepsilon_1{+}\varepsilon_2)^2\right]b^{\dagger}_{\varepsilon_1 \rm r} b_{\varepsilon_2 \rm r},
\nonumber\\
\mathscr{H}_{\rm int}&{=}&\sum_{\rm r < r'}\int_{\varepsilon_{1-4}}\left[\frac{\phi_1}{\pi\nu}{+}\frac{\phi_2}{4\pi\nu}\sum^4_{\rm j=1}\varepsilon_{\rm j}\right]{:}b^{\dagger}_{\varepsilon_1 \rm r} b_{\varepsilon_2 \rm r} b^{\dagger}_{\varepsilon_3 \rm r'} b_{\varepsilon_4 \rm r'}{:}.\nonumber
\end{eqnarray}
Here the spin index $r$ takes all the possible values from 1 to $N$. Note that there is no zeroth order process in elastic part of the Hamiltonian $\mathscr{H}_{\rm el}$ since it describes merely the ground state property namely the energy dependent scattering processes in FL. However, since the interaction (the property of excitation) among the FL quasi particles is accounted for by the Hamiltonian $\mathscr{H}_{\rm int}$, it does starts from the zeroth order term.

The scattering phase shift described by above Hamiltonian reads~\cite{mora1, mora2}
\begin{align}\label{manb0}
&\delta_{\rm r}(\varepsilon)=\delta_0+\alpha_1\varepsilon+\alpha_2\varepsilon^2-\sum_{\rm r'\neq r}\Bigg[\phi_1\int^{\infty}_{-\infty}d\varepsilon \delta n_{\rm r'}(\varepsilon)\nonumber\\
&+\frac{\phi_2}{2}\left(\varepsilon\int^{\infty}_{-\infty}d\varepsilon \delta n_{\rm r'}(\varepsilon)+\int^{\infty}_{-\infty}d\varepsilon \varepsilon \delta n_{\rm r'}(\varepsilon)\right)\Bigg].
\end{align}
Here the zeroth order processes in energy are controlled by the scattering phase shift
\begin{equation}
\delta_0=m\pi/N.
\end{equation}
The symbol $\delta n_{\rm r}(\varepsilon)$ stands for the actual FL quasi-particle distribution relative to the Fermi-energy $\varepsilon_{\rm F}$ 
 \begin{equation}
 \delta n_{\rm r}(\varepsilon)\equiv n_{\rm r}(\varepsilon)-{\rm \Theta}(\varepsilon_{\rm F}-\varepsilon)=\langle b^{\dagger}_{k\rm r}b_{k\rm r}\rangle-{\rm \Theta}(\varepsilon_{\rm F}-\varepsilon),
 \end{equation}
where $\Theta$ is the step function and the average $\langle b^{\dagger}_{k\rm r}b_{k\rm r}\rangle$ is expressed in terms of the equilibrium Fermi-distribution functions of the left and right reservoirs such that $\langle b^{\dagger}_{k\rm r}b_{k\rm r}\rangle=\left(f_{\rm L}+f_{\rm R}\right)/2$. Using the Kondo-floating ansatz~\cite{mora1, mora2}, we then chose the Fermi-level satisfying the condition
\begin{equation}\label{kaule2}
\int^{\infty}_{-\infty}d\varepsilon\delta n_{\rm r}(\varepsilon)=0.
\end{equation}
Using all the specifications as described above, the simple energy integration of Eq.~\eqref{manb0} leads an expression of the equilibrium phase shift
\begin{align}
\label{phase_shift}
&\delta_{\rm r} (\varepsilon)=\delta_0+\alpha_1\varepsilon+\alpha_2\varepsilon^2- \frac{(N-1)}{12}\phi_2(\pi T)^2.
\end{align}

From the scattering phase shift Eq.~\eqref{phase_shift}, we get an expression of the total T-matrix~\citep{GP_Review_2005}
\begin{equation}\label{manbd2}
-\pi\nu \mathcal{T}^{\rm tot}_{\rm r}(\varepsilon)=\frac{1}{2i}\left[e^{2i\delta_{\rm r}(\varepsilon)}-1\right]+e^{2i\delta_{\rm r}(\varepsilon)}\left[-\pi\nu \mathcal{T}^{\rm in}_{\rm r}(\varepsilon)\right],
\end{equation}
where the T-matrix describing the FL quasi-particle interactions have the usual form~\cite{GP_Review_2005,mora1, mora2}
\begin{equation}
-\pi\nu \mathcal{T}^{\rm in}_{\rm r}(\varepsilon)=-\frac{N-1}{2i}\left[\varepsilon^2+(\pi T)^2\right]\phi_1^2.
\end{equation}
From Eq.~\eqref{manbd2} we extract the imaginary and real part of the T-matrix, which read for the imaginary part
\begin{align}\label{manbd3}
&{\rm Im}\left[-\pi\nu \mathcal{T}^{\rm tot}_{\rm r}(\varepsilon, T)\right]=\sin^2\delta_0+\frac{1}{2}(\pi T)^2 (N-1)\nonumber\\
&\;\;\;\;\times\left(\phi^2_1  \cos2\delta_0-\frac{1}{6}\phi_2\sin2\delta_0\right)+ \alpha_1 \sin2\delta_0\;\varepsilon\nonumber\\
&\;\;\;\;+\left[\alpha_2\sin2\delta_0 + \cos2\delta_0\left(\alpha^2_1 +\frac{N-1}{2}\phi_1^2\right)\right]\varepsilon^2,
\end{align}
and for the real part we have
\begin{align}\label{manbd4}
&{\rm Re}\left[-\pi\nu \mathcal{T}^{\rm toto}_{\rm r}(\varepsilon, T)\right]=\frac{1}{2}\sin2\delta_0-\frac{1}{12} (\pi T)^2 (N-1)\nonumber\\
&\;\;\;\;\;\times\left[\phi_2\cos2\delta_0+6\phi^2_1\sin2\delta_0\right]+ \alpha_1  \cos2\delta_0\;\varepsilon\nonumber\\
&\;\;\;\;\;+\Big[ \alpha_2 \cos2\delta_0-  \sin2\delta_0\left(\alpha_1^2+\frac{N-1}{2}\phi^2_1\right)\Big]\varepsilon^2.
\end{align}

Having the expressions of real and imaginary parts of the T-matrix in Eqs.~\eqref{manbd3} and~\eqref{manbd4}, we first obtained their thermal and spin averaged values as defined in Eq.~\eqref{sita2} by using the Sommerfeld integrals
\begin{equation}\nonumber
\mathcal{I}_n\equiv\frac{1}{4 T}\int^{\infty}_{-\infty}d\varepsilon\frac{\varepsilon^n}{\cosh ^2\left(\frac{\varepsilon}{2 T}\right)},\;\;\mathcal{I}_0=1,\;\;\mathcal{I}_2=\frac{1}{3}(\pi T)^2.
\end{equation}
Our results for the thermal and spin averaged values of the imaginary and real part of the T-matrix are given by
\begin{align}\label{manbd5}
\pi\;{\rm Im}\mathcal{J}=\sin^2\delta_0&+\frac{(\pi T)^2}{3}\Big[\sin2\delta_0\Big(\alpha_2-\frac{N-1}{4}\phi_2\Big)\nonumber\\
&+\cos2\delta_0\Big(\alpha^2_1+2(N-1)\phi^2_1\Big)\Big],
\end{align}
and 
\begin{align}\label{manbd6}
\pi\;{\rm Re}\mathcal{J}=\frac{1}{2}\sin2\delta_0&+\frac{(\pi T)^2}{3}\Big[\cos2\delta_0\Big(\alpha_2-\frac{N-1}{4}\phi_2\Big)\nonumber\\
&-\sin2\delta_0\Big(\alpha^2_1+2(N-1)\phi^2_1\Big)\Big].
\end{align}
The TPS then directly follows from Eqs.~\eqref{manbd5} and~\eqref{manbd6} with the definition $\theta_{\rm tp}{=}\arctan\left[{\rm Im}\mathcal{J}/{\rm Re}\mathcal{J}\right]$ as presented in Eq.~\eqref{sita2}. For completeness, we explicitly write the mathematical expression of TPS
\begin{widetext}
\begin{equation}\label{manbd7}
\theta_{\rm tp}=\tan^{-1}\Bigg(\frac{\sin^2\delta_0+\frac{(\pi T)^2}{3}\Big[\sin2\delta_0\Big(\alpha_2-\frac{N-1}{4}\phi_2\Big)
+\cos2\delta_0\Big(\alpha^2_1+2(N-1)\phi^2_1\Big)\Big]}{\frac{1}{2}\sin2\delta_0+\frac{(\pi T)^2}{3}\Big[\cos2\delta_0\Big(\alpha_2-\frac{N-1}{4}\phi_2\Big)
-\sin2\delta_0\Big(\alpha^2_1+2(N-1)\phi^2_1\Big)\Big]}\Bigg).
\end{equation}
\end{widetext}
In the following subsections we discuss the TPS evolution in 1CK effects separately for different cases.
\subsection{TPS and normalized visibility with exact SU($N$) symmetry}
In the limit of exact $SU(N)$ symmetry various exact identities connection different FL coefficients are available. The connection of first and second generation of FL coefficients, $\alpha_1$ and $\alpha_2$ respectively, characterizing the elastic effects is given by~\cite{Nozieres, mora1}
\begin{equation}\label{manbd8}
\frac{\alpha_2}{\alpha^2_1}=\frac{N-2}{N-1}\frac{\Gamma(1/N)\tan(\pi/N)}{\sqrt{\pi}\Gamma\left(\frac{1}{2}+\frac{1}{N}\right)}\cot\left[\frac{m\pi}{N}\right].
\end{equation}
Here the symbol $\Gamma(x)$ defines the Euler's gamma-function. Note that for the half-filled systems, $m{=}N/2$, the second generation of the FL-coefficient gets exactly nullified. In addition, the connections between the elastic FL-coefficients and those of inelastic one read~\cite{mora2}
\begin{equation}\label{manbd9}
\alpha_1=(N-1)\phi_1,\;\;\;\alpha_2=\frac{N-1}{4}\phi_2.
\end{equation}
In addition, without loss of generality we define the Kondo temperature of the $SU(N)$ system by simple scaling relation
\begin{equation}\label{manbd10}
\Tk\equiv\frac{1}{\alpha}_1.
\end{equation}

From Eqs.~\eqref{manbd8},~\eqref{manbd9} and~\eqref{manbd10}, it is apparent that the low energy sector of $SU(N)$ Kondo effects is fully described by the single energy scale $\alpha_1$ or equivalently the Kondo temperature $\Tk$. By plugging in the Eqs.~\eqref{manbd8},~\eqref{manbd9} and~\eqref{manbd10} into Eq~\eqref{manbd7} followed by the simple mathematical simplification, we obtained the expression of TPS, whose low temperature expansion reads
\begin{equation}\label{manbd11}
\left.\theta_{\rm tp}\right|_{m, N}=\frac{m\pi}{N}+\mathscr{C}_{{\rm T},N}(m) \left(\frac{\pi T}{\Tk}\right)^2,
\end{equation}
with the coefficient $\mathscr{C}_{{\rm T},N}(m)$ characterizing the finite temperature correction
\begin{equation}
\mathscr{C}_{{\rm T},N}(m)=\frac{1}{3}\frac{N+1}{N-1}\cot \left(\frac{\pi  m}{N}\right).
\end{equation}

From the last Eq.~\eqref{manbd11} it is seen that the coefficient $\mathscr{C}_{{\rm T}, N}(m)$ can be considered as distinctive FL coefficient characterizing SU($N$) Kondo effects. Once $m$ is changed to even value keeping $N$ also even this numbers ultimately vanishes owing to have the exact PH symmetric situation where the $\pi/2$ phase locked has been experimentally observed. For odd $N$, however, with any given numbers of electrons $m$ the PH symmetric analog doesn't simply exists and hence the TPS acquires the finite temperature correction. This finite temperature corrections in $SU(N)$ Kondo effects are likely to be observed with the existing experimental setup~\cite{delft_andrea}.

For case of $N{=}2m$ the small PH symmetry breaking effects can be accounted for by considering the potential scattering phase shift $\delta_{\rm P}$ such that $\delta_0=\frac{\pi}{2}\pm\delta_{\rm p}$ for $\delta_{\rm P}\ll\delta_0$~\cite{Affleck_Lud_PRB(48)_1993}
\begin{equation}\label{manbd12}
\mathscr{C}_{{\rm T},N}(m=N/2)=\mp\frac{1}{3}\frac{N+1}{N-1}\tan\delta_{\rm p}.
\end{equation}
From above discussion it is instructive to think of the coefficient $\mathscr{C}_{{\rm T},N}(m)$ as the distinctive FL constant of generic SU($N$) Kondo impurity, which could be of paramount importance for ongoing transport experiments in Kondo correlated systems. The Eq.~\eqref{manbd12} provides the qualitative idea of TPS evolution beyond PH symmetric point of the $SU(N{=}2m)$ Kondo effects. In the following section, we present the microscopic calculations starting from the single level Anderson model for TPS in paradigmatic $SU(2)$ Kondo effects beyond PH symmetric point.

From Eqs.~\eqref{manbd5} and~\eqref{manbd6} we get the expression for the normalized visibility in $SU(N)$ Kondo regime
\begin{equation}\label{aar1}
\left.\eta\right|_{m, N}=1-\frac{1}{3\sin^2\left(\frac{m\pi}{N}\right)}\;\frac{N+1}{N-1}\;\left(\frac{\pi T}{\Tk}\right)^2.
\end{equation}
At the zero temperature case of fully screened Kondo effects, the formation of Kondo singlet implies that all the particles that are scattered off the impurity 
are scattered into single particle states. Consequently, all the conductance is carried by the coherent processes. At finite temperature the phase shift acquires the energy dependence and also the inelastic effects become prominent. Both of these processes result in the finite temperature reduction of visibility as expressed by Eq.~\eqref{aar1}.

\subsection{TPS evolution beyond PH symmetric point in paradigmatic $SU(2)$ Kondo effects}
The $SU(2)$ Kondo effects are one of the most intensely studied variant of $SU(N)$ framework. From the previous section, it is seen that the PH symmetric point result in the $\pi/2$ phase shift between outgoing and incoming electron states. However, the PH symmetric description is not always sufficient to analyze the real experiment. The single-impurity Anderson model~\cite{and1} provides an straightforward way of proceeding with the qualitative description for the transport features beyond PH symmetric point of the $SU(2)$ Kondo correlated systems.

The second generation of the FL coefficients $\alpha_2$ and $\phi_2$ (see previous section) incidentally vanish owing to have the PH symmetry. Therefore, the original theory of 1CK proposed by Nozieres~\cite{Nozieres} contains two FL parameters namely $\alpha_1$ and $\phi_1$. Beyond PH symmetric transport description, thus, involves the four FL parameters $\alpha_{1, 2}$ and $\phi_{1, 2}$ in addition the zero energy phase shift $\delta_0$~\cite{jvm}. The general idea to proceed further would then be to make the connection of five FL parameters with the original parameters of the Anderson model. The generic single-impurity Anderson model is described by three parameters: the impurity energy level $\ed$, level hybridization $\mathcal{F}$ and the Coulomb energy $U$~\cite{and1}.

The connection between the two set of parameters, the FL parameters and parameters of Anderson model, for some limiting cases has been investigated in the recent work~\cite{jvm}. The emergent of Kondo regime can be assured with fulfilling the condition $U\gg\mathcal{F}$ and $-U+\mathcal{F}<\ed<-\mathcal{F}$. The latter condition can be polished to express it into the more symmetrical form
\begin{equation}\label{manbd13}
-\frac{1}{2}\left(\frac{U}{\mathcal{F}}-2\right)<\frac{1}{\mathcal{F}}\left(\ed+\frac{U}{2}\right)<\frac{1}{2}\left(\frac{U}{\mathcal{F}}-2\right).
\end{equation}
In this limit, FL-coefficients can be expressed merely in terms of the spin susceptibility $\chi_{\rm s}$ and its derivative $\chi'_{\rm s}$ with respect to $\ed$~\cite{jvm} 
\begin{equation}\label{manbd14}
\alpha_1\simeq\phi_1\simeq\pi\chi_{\rm s},\;\;\alpha_2=\frac{3}{4}\phi_2\simeq-\pi\chi'_{\rm s}.
\end{equation}
The Eq.~\eqref{manbd14} provides the $\ed$, $\mathcal{F}$ and $U$ dependence of FL coefficients $\alpha_{1, 2}$ since the spin susceptibility $\chi_{\rm s}$ can elegantly be expressed in terms of the original parameter of the Anderson model~\cite{jvm1}. The actual connections are given by~\cite{jvm}
\begin{equation}\label{manbd16}
\oa_1=\frac{\pi}{2\sqrt{2}}\frac{1}{\sqrt{\mb}}\exp\left[\pi\left(\frac{\mb}{8}-\frac{1}{2\mb}\right)-\frac{\pi}{2}\frac{\ma^2}{\mb}\right],
\end{equation}
\begin{equation}\label{manbd17}
\oa_2=\frac{\pi^2}{2\sqrt{2}}\frac{\ma}{\mb^{3/2}}\exp\left[\pi\left(\frac{\mb}{8}-\frac{1}{2\mb}\right)-\frac{\pi}{2}\frac{\ma^2}{\mb}\right],
\end{equation}
with the re-scaled parameters
\begin{equation}\nonumber
\oa_1=\mathcal{F}\alpha_1,\;\;\oa_2=\mathcal{F}^2\alpha_2,\;\;\ma=\frac{1}{\mathcal{F}}\left(\ed{+}\frac{U}{2}\right)\!,\;\;\mb=\frac{U}{\mathcal{F}}.
\end{equation}

Substitution of the Eq.~\eqref{manbd14} into Eq.~\eqref{manbd7} for $N{=}2$ results in the TPS for $SU(2)$ Kondo effects beyond PH symmetric point
\begin{equation}\label{aar2}
\tan\theta_{\rm tp}=\frac{\sin^2\delta_0+\left(\pi\ot\right)^2 \Big(\overline{\alpha}^2_1  \cos2\delta_0
+\frac{2}{9}\overline{\alpha}_2\sin2\delta_0\Big)}{\frac{1}{2}\sin2\delta_0-\left(\pi\ot\right)^2 \Big(\overline{\alpha}^2_1  \sin2\delta_0
-\frac{2}{9}\overline{\alpha}_2\cos2\delta_0\Big)},
\end{equation}
with the re-scaled temperature $\ot{=}T/\mathcal{F}$. The lowest order temperature correction for TPS then reads
\begin{equation}\label{manbd15}
\theta_{\rm tp}(\mathscr{A}, \mathscr{B})=\delta_0+ \left(\pi\ot\right)^2\left[\frac{2}{9}\;\overline{\alpha}_2+\overline{\alpha}^2_1\cot\delta_0\right].
\end{equation}
Therefore the finite temperature TPS depends on three parameters $\delta_0$, $\overline{\alpha}_1$ and $\overline{\alpha}_2$. The connections of $\overline{\alpha}_1$ and $\overline{\alpha}_2$ to the original parameters of the Anderson model $\mathscr{A}$ and $\mathscr{B}$ are provided in Eqs.~\eqref{manbd16} and~\eqref{manbd17} respectively. The remaining parameter, the zero energy phase shift, further depends on $\ma$ and $\mb$ in rather non-trivial way, which can be obtained by solving the coupled equations~\cite{jvm1, jvm}
\begin{align}
\ma= & \Theta(q)\sqrt{2\mb q}-\frac{\sqrt{\mb}}{2\pi^{3/2}}\times{\rm Re}\Bigg[\frac{1}{\sqrt{i}}\int^{\infty}_0 dy\;\frac{e^{-2iq\pi y}}{y^{3/2}}\times\nonumber\\
& \Bigg\{e^{-\pi y}\left(\frac{e}{iy}\right)^{iy}\Gamma\left(\frac{1}{2}+iy\right)-\sqrt{\pi}\Bigg\}\Bigg],
\end{align}
and
\begin{align}
&\delta_0(\ma; \mb)= \frac{\pi}{2}\Bigg(\frac{1}{2}-\frac{1}{\pi^{3/2}}\times{\rm Re}\Big[i\int^{\infty}_0 dy\;\frac{e^{-2iq\pi y}}{y}e^{-\pi y}\nonumber\\
&\times\left(\frac{e}{iy}\right)^{iy}\Gamma\left(\frac{1}{2}+iy\right)\int^{\infty}_{-\infty}\frac{dx}{\pi}\frac{e^{i\pi y x^2/\mb}}{1+\left(x+\frac{\mb}{2}\right)^2}\Big]\Bigg).
\end{align}
In the last equations, $\Theta$ stands for the unit step function and $q$ is an arbitrary parameter. With all the specifications discussed above, the Eq.~\eqref{manbd15} provides the generic feature of finite temperature TPS evolution at and beyond the PH symmetric point for paradigmatic $SU(2)$ Kondo correlated systems.

Before closing this subsection, we briefly discuss about the dependence of the normalized visibility on the microscopic parameters $\ed$, $\mathcal{F}$ and $U$. Although the required relation can be directly read out from the Eq.~\eqref{aar2}, the more straightforward way of getting compact expression is to use the Eq.~\eqref{aar1} for $N{=}2$ in terms original parameters of the Anderson model which reads
\begin{equation}
\eta(\mathscr{A}, \mathscr{B})=1-\frac{\overline{\alpha}_1^2 }{\sin^2\delta_0}\left(\frac{\pi T}{\mathcal{F}}\right)^2.
\end{equation}
\section{Overscreened Kondo effects}\label{sec3}
The ground state of a quantum impurity with an effective spin $\mathcal{S}$ in the overscreened Kondo regime $2\mathcal{S}<\mathcal{K}$ have a finite residual impurity spin of magnitude $|\mathcal{S}-\mathcal{K}/2|$. The situation for $\mathcal{K}$CK is, therefore, strikingly different as compared to the conventional 1CK. The zero temperature limit for conventional 1CK is completely described by the single particle scattering off the impurity~\cite{Nozieres}. Nevertheless, the finite probability for the multi particle scattering off a impurity at zero temperature is one of the hallmark of overscreened Kondo effects~\cite{Nozieres_Blandin_JPhys_1980}, which signifies the NFL correlation. This characteristics of an overscreened $\mathcal{K}$CK results in the dramatic change of transport features over the corresponding 1CK effects. The common example includes the temperature dependence of the differential conductance acquires the fractional critical exponent that depends on the number of screening channels $\mathcal{K}$~\cite{Affleck_Lud_PRB(48)_1993}. In this section, we investigate the influences of NFL correlation on the TPS behavior of generic overscreened $\mathcal{K}$CK using the seminal work of AL based on the BCFT~\cite{Affleck_Lud_PRB(48)_1993}.
\subsection{Formulation of T-matrix}
We consider an spin-1/2 overscreened Kondo impurity with $\mathcal{K}$ conduction channels in the AL framework. In order to account for the effects of PH symmetry breaking, we explicitly consider the potential scattering on the top of Kondo interaction. The potential scattering is accounted for by considering the phase shift $\delta_{\rm P}$. The T-matrix of $\mathcal{K}$CK effects then reads~\cite{Affleck_Lud_PRB(48)_1993}
\begin{align}\label{kck1}
-\pi\nu\mt(\varepsilon, T) &=\frac{i}{2}\Big[\left(1-e^{2i\delta_{\rm P}}\rss\right)-e^{2i\delta_{\rm P}} Z \;\mathcal{I}\Big].
\end{align}
In Eq~\eqref{kck1}, the symbol $\rss$ stands for the S-matrix in the single particle sector of $\mathcal{K}$CK effects. The S-matrix is the function of the spin of the impurity $\mathcal{S}$, spin of conduction electrons $j$ and number of conduction channels $\mathcal{K}$~\cite{Affleck_Lud_PRB(48)_1993}
\begin{equation}\label{yemadd}
\rss=\frac{\sin \left(\frac{\pi }{\mathcal{K}+2}\right)  \sin \left(\frac{\pi  (2 j+1) (2 \mathcal{S}+1)}{\mathcal{K}+2}\right)}{\sin \left(\frac{\pi(2  j+1) }{\mathcal{K}+2}\right) \sin\left(\frac{ \pi(2  \mathcal{S}+1) }{\mathcal{K}+2}\right)}.
\end{equation}
In the following, we focus on the particular case of $j, \mathcal{S}=1/2$ and denote the S-matrix as
\begin{equation}
\left.\rss\right|_{j, \mathcal{S}=1/2}\equiv\rss=\cos \left(\frac{ 2\pi}{\mathcal{K}+2}\right)\Big/
\cos \left(\frac{\pi}{\mathcal{K}+2}\right),
\end{equation}
and the general case of arbitrary impurity spin and spin of conduction electrons will be treated later. 
The low energy physics of the problem is fundamentally controlled by the parameter $Z$ which is defined in terms of the leading irrelevant operator (LIO) $\lambda$, number of orbital channels $\mathcal{K}$ and the temperature $T$ such that~\cite{Affleck_Lud_PRB(48)_1993}
\begin{align}\label{kto1}
Z=\mathscr{R}(\Delta)\lambda(\pi T)^{\Delta},\;\;\;\Delta\equiv \frac{2}{\mathcal{K}+2},
\end{align}
with the parameter $\mathscr{R}$ defined by the relation
\begin{equation}\nonumber
\mathscr{R}(\Delta)=3{\times} 2^{\Delta}\sin\left(\pi\Delta\right){\sqrt{\!\frac{{\sin \left(\frac{\pi }{\mathcal{K}{+}2}\right)}{ \tan \left(\frac{\pi }{\mathcal{K}{+}2}\right)}{ \Gamma \left(\frac{\mathcal{K}}{\mathcal{K}{+}2}\right)^2}}{\Gamma \left(\frac{\mathcal{K}{-}1}{\mathcal{K}{+}2}\right)\Gamma \left(\frac{\mathcal{K}{+}1}{\mathcal{K}{+}2}\right)}}}.
\end{equation}
In addition the energy and temperature dependent parameter $\mathcal{I}$ in Eq~\eqref{kck1} reads
\begin{equation}
\mathcal{I}\equiv \int^{1}_{0} du\left[ \frac{u^{-\frac{i\varepsilon}{2\pi T}}(1-u)^{\Delta}}{\sqrt{u}}F(u, \Delta)-\mathscr{X}(u, \Delta)\right],
\end{equation}
with the new definition
\begin{equation}
\mathscr{X}(u, \Delta)=\frac{\Gamma[1+2\Delta]}{\Gamma^2[1+\Delta]}u^{\Delta-1}(1-u)^{-(1+\Delta)},
\end{equation}
and the hypergeometric functions $_2F_1(1+\Delta, 1+\Delta; 1; u)\equiv F(u, \Delta)$
\begin{equation}
F(u, \Delta)=\frac{1}{2\pi}\int^{2\pi}_{0}\frac{d\theta}{(u+1-2\sqrt{u}\cos\theta)^{1+\Delta}}.
\end{equation}
We then extract the real and imaginary parts of the T-matrix from Eq~\eqref{kck1}, which cast into the compact form
\begin{align}
{\rm Im}\left(-\pi\nu\mt\right)&{=}\frac{1}{2}\left(1-\rss\cp\right)\left[1-Q_2\mathcal{I}_1-Q_3\mathcal{I}_2\right],\nonumber\\
{\rm Re}\left(-\pi\nu\mt\right)&{=}\frac{1}{2}\left(1{-}\rss\cp\right)\left[Q_1{-}Q_2\mathcal{I}_2{+}Q_3\mathcal{I}_1\right].\label{kck2}
\end{align}
Here we defined the new symbols $Q_{j}$ 
\begin{align}
Q_1=&\frac{\rss\;\cs}{1-\cp\rss},\\
Q_2=&\frac{Z\;\cp}{1-\cp\rss},\\
Q_3=&\frac{Z\;\cs }{1-\cp\rss}.
\end{align}
In addition, the short-hand notations $\mathcal{I}_{1, 2}$ in Eq~\eqref{kck1} depend on the temperature and energy for given value of $\Delta$ in such a way that
\begin{align}
\mathcal{I}_1&{\equiv}\!\int^{1}_{0}\!\!\! du\left[\!\cos\left[\!\frac{\varepsilon}{2\pi T}\!\log(u) \right]\!\!\frac{(1{-}u)^{\Delta} F(u, \Delta)}{\sqrt{u}}{-}\mathscr{X}(u, \Delta)\right],\nonumber\\
\mathcal{I}_2&{\equiv} \int^{1}_{0}\!\!\!du\left[\sin\left[\frac{\varepsilon}{2\pi T}\log(u) \right]\frac{(1-u)^{\Delta} F(u, \Delta)}{\sqrt{u}}\right].
\end{align}
\subsection{TPS evolution in overscreened $\mathcal{K}$CK effects}
In this subsection we proceed for the calculation of the thermally averaged values of the real and imaginary parts of the T-matrix needed for computing the the TPS for $\mathcal{K}$CK effects. To this end, the Eq.~\eqref{kck1} provides the imaginary part of thermally averaged value of T-matrix
\begin{align}\label{im1}
{\rm Im}\mathcal{J}&= \frac{1}{4T}\sum_{\rm r=\uparrow,\downarrow}\int^{\infty}_{-\infty}\frac{d\varepsilon}{2\pi}\;\frac{{\rm Im}\left[-\pi\nu\mathcal{T}(\varepsilon, T)\right]}{\cosh^2(\varepsilon/2T)}\nonumber\\
&\equiv \frac{1}{2\pi}\left(1-\rss\cp\right)\left(1-B_1-B_2\right).
\end{align}
To obtain Eq~\eqref{im1} we changed the order of integration between the energy $\varepsilon$ and the variable $u$ followed by defining the energy integrals $B_{1, 2}$ in terms of $\mathcal{I}_{1, 2}$
\begin{align}
B_j & \equiv \frac{Q_{j+1}}{4T}\int^{\infty}_{-\infty}d\varepsilon
\frac{\mathcal{I}_j}{\cosh^2(\varepsilon/2T)},\;j=1, 2.
\end{align}
By symmetry the integral $B_2$ vanishes while $B_1$ simplifies to
\begin{align}
B_1=-Q_2\mathscr{P}(\Delta),
\end{align}
with the new definition
\begin{equation}
\mathscr{P}(\Delta)\equiv\int^{1}_{0}du\left[\frac{F(u)\log(u)}{(1-u)^{1-\Delta}}+\mathscr{X}(u, \Delta)\right].
\end{equation}
To arrived at the last equation the exact integral related to the Fourier-transform of Fermi-function has been employed
\begin{equation}\nonumber
\int^{\infty}_{-\infty}\frac{d\varepsilon}{\cosh^2(\varepsilon/2T)}\cos\left[\frac{\varepsilon}{2\pi T}\ln u\right]=\frac{4 T \sqrt{u} \log (u)}{u-1}.
\end{equation}
This completes the calculation of the thermally averaged value for the imaginary part of the T-matrix. Similar procedures have then be used for the calculation of the thermally averaged real part of the T-matrix. To list the results, we have
\begin{align}
\pi\;{\rm Im}\mathcal{J}&=\frac{1}{2}\Big[1-\rss\cp+Z\cp\mathscr{P}(\Delta)\Big],\label{kto3}\\
\pi\;{\rm Re}\mathcal{J}&=\frac{1}{2}\Big[ \rss\cs-Z\cs\mathscr{P}(\Delta)\Big].\label{kto4}
\end{align}

The LIO $\lambda$ in the expression of $Z$ Eq.~\eqref{kto1} depends on the Kondo temperature $\Tk$. In this work, without loss of generality, we define the Kondo temperature as given by
\begin{equation}\label{kto2}
\lambda\equiv-\left(\frac{1}{\Tk}\right)^{\Delta}.
\end{equation}
By plugging in the Eqs.~\eqref{kto3}-~\eqref{kto2} into Eq.~\eqref{sita2}, we obtain a compact form of TPS in overscreened $\mathcal{K}$CK regime
\begin{equation}\label{kto5}
\theta_{\rm tp}{=}{\tan^{-1}}\!\Bigg[\!\frac{1{-}\cp \frac{\cos \left(\pi\Delta\right)}{
\cos \left(\pi\Delta/2\right)}{-}\cp\mathscr{D}(\Delta)\!\left(\frac{\pi T}{\Tk}\right)^{\Delta}}{\cs\frac{\cos \left(\pi\Delta\right)}{
\cos \left(\pi\Delta/2\right)}+\cs\;\mathscr{D}(\Delta)\!\left(\frac{\pi T}{\Tk}\right)^{\Delta}}\!\!\Bigg],
\end{equation}
with the new definition
\begin{equation}
\mathscr{D}(\Delta)=\mathscr{R}(\Delta)\;\mathscr{P}(\Delta).
\end{equation}
The Eq.~\eqref{kto5} is the general expression providing the TPS evolution in overscreened Kondo regime. The function $\mathscr{P}(\Delta)$ has recently been computed to be~\cite{dbk}
\begin{equation}\label{ktt1}
\mathscr{P}(\Delta)=\frac{1}{\Delta (1+\Delta)}.
\end{equation}
The Eqs.~\eqref{kto1},~\eqref{kto5} and~\eqref{ktt1} provide the complete evolution of TPS in overscreened $\mathcal{K}$CK effects.

Of the particular experimental interest is the 2CK, where the single-particle $\rss$-matrix vanishes resulting in the absence of single-particle scattering off a impurity at zero temperature. For this special case of 2CK, $\mathscr{D}(\Delta){=}4$ and $\Delta{=}1/2$ resulting in the greatly simplified form of TPS
\begin{equation}\label{kto6}
\theta^{\rm 2CK}_{\rm tp}={\tan^{-1}}\!\Bigg[\!\frac{1-4\cp\sqrt{\pi T/\Tk}}{4\cs\sqrt{\pi T/\Tk}}\Bigg].
\end{equation}
From the last Eq.~\eqref{kto6}, we recover the $\pi/2$ phase shift at the PH symmetric point ($\delta_{\rm P}{=}0$). Beyond PH symmetric point, however, the TPS acquires the significant temperature correction from the corresponding zero temperature value of $\pi/2$. In addition the normalized visibility can be easily read out from the Eqs.~\eqref{kto3} and~\eqref{kto4}.

In the previous subsections we considered the particular case of spin-1/2 impurity screened by $\mathcal{K}$ channels of spin-1/2 conduction electrons. Therefore, before closing this subsection we briefly discuss the coherent properties of overscreened Kondo effects with an arbitrary spin $\mathcal{S}$. For the general spin amplitude $\mathcal{S}$ the treatment follows similarly as per the spin-1/2 situation presented earlier. However, the spin $\mathcal{S}$ affects the general behavior of single-particle scattering matrix such that [see Eq.~\eqref{yemadd} with $j=1/2$]
\begin{equation}\label{kto7}
\rss=\cos \left[\frac{ \pi  (2 \mathcal{S}+1)}{\mathcal{K}+2}\right]\Big/
\cos \left[\frac{\pi}{\mathcal{K}+2}\right].
\end{equation}
In addition, consideration of general $\mathcal{S}$ changes the expression of $Z$ in Eq.~\eqref{kto1} by its explicit dependence on the factor $\mathscr{R}(\Delta)$
\begin{equation}
\mathscr{R}(\Delta)=2^{\Delta+1}\mathcal{X}\;\sin(\pi\Delta),
\end{equation}
with the numerical factor~\cite{Affleck_Lud_PRB(48)_1993}
\begin{align}
\mathcal{X}\equiv& \Bigg[\frac{\Gamma \left[\frac{\mathcal{K}}{\mathcal{K}+2}\right]^2}{\Gamma \left[\frac{\mathcal{K}+1}{\mathcal{K}+2}\right] \Gamma \left[\frac{\mathcal{K}-1}{\mathcal{K}+2}\right]}\frac{\cos \left[\frac{2 \pi }{\mathcal{K}+2}\right]\;\sec \left[\frac{ \pi }{\mathcal{K}+2}\right]-\rss}{1+2 \cos \left[\frac{2 \pi }{\mathcal{K}+2}\right]}\Bigg]^{\frac{1}{2}}.
\end{align}

Of the particular interest, for this work, are the situation with $\mathcal{K}{=}4\mathcal{S}$. As seen from the Eq.~\eqref{kto7} that the special combinations $\mathcal{K}{=}4\mathcal{S}$ have the vanishing single-particle scattering matrix $\rss$ therefore the complete absence of single particle to single particle scattering processes. The most general cases reflecting this peculiar behavior are the two channel spin-1/2 Kondo effects. In general, the cases with $\mathcal{K}=2, 4, 6\cdots$ with spin combination $\mathcal{S}=1/2, 1,3/2\cdots$ satisfy the exact nullification of single-particle scattering matrix. At PH symmetric point, the TPS is always $\pi/2$ free from the spin amplitude $\mathcal{S}$. The normalized visibility at the lowest order of temperature (with $\delta_{\rm P}=0$) for these special cases reads
\begin{eqnarray}\label{kappa3}
\eta^{\mathcal{K}=\!4\mathcal{S}}{=}\frac{1}{2}\Big[1{-}2^{\Delta{+}1}\mathcal{X}\sin(\pi\Delta)\mathscr{P}(\Delta)\!\left(\pi T/\Tk\right)^{\Delta}\!\Big].
\end{eqnarray}
Although the scattering amplitude $\rss$ gets vanishes, the finite temperature behavior of $\eta$ therefore depends on the $\mathcal{S}$ via the coefficient in front of the temperature in Eq.~\eqref{kappa3}. 

It is also worth of mentioning that, the exact vanishing of the single-particle to single-particle scattering amplitude can also be realized with the recent proposal exhibiting the topological Kondo effect~\cite{beri}. The minimal setup for the topological Kondo effect corresponds to the case of $j=1$, $\mathcal{S}=1/2$ and $\mathcal{K}=4$ in Eq.~\eqref{yemadd} which results in $\rss=0$. In addition, all the results and discussion presented in our work can straightforwardly be generalized for the corresponding descriptions in topological Kondo regime. 

Before closing this subsection, we want to stress that the definition of LIO $\lambda$ in terms of the Kondo temperature $\Tk$ depends on the initial coupling strength. Although, we defined the kondo temperature via the relation $\lambda=-\left(1/\Tk\right)^{\Delta}$, the corresponding definition with the positive sign $\lambda=\left(1/\Tk\right)^{\Delta}$ would be equally applicable~\cite{carmi1}.
\section{Conclusion}\label{sec4}
We develop a full-fledged theoretical framework to study the coherent properties of fully screened and overscreened Kondo effects by exploiting the AB ring. The developed formalism is equally applicable for the transport description at and beyond the PH symmetric point and therefore provides the complete evolution of TPS and normalized visibility in Kondo correlated systems. In the realm of 1CK, we extended the commonly studied TPS behavior in the conventional $SU(2)$ Kondo regime to that in a high-symmetry $SU(N)$ Kondo paradigm. We explored a distinctive FL coefficient characterizing the finite temperature correction to the TPS and normalized visibility that can be measured with the existing experimental setup~\cite{delft_andrea}. In addition, we provided a general mathematical framework to deal with the TPS behavior beyond the PH symmetric point in the conventional $SU(2)$ Kondo regime. Our work equally provides the complete picture of TPS evolution in overscreened $\mathcal{K}$CK effects. In particular, the constructed theoretical framework based on boundary conformal-field-theory shows the crucial role of finite temperature effects on the coherent properties in the NFL regime associated with the overscreened Kondo effects. In addition, the expression of the T matrices explored in this work could be equally useful for the study of thermoelectric transport through the Kondo impurities both in the FL and NFL regime. Although we focused our discussion on the original spin Kondo effects, the developments in this work are equally applicable for the exotic topological Kondo effects. In this work, we used the BCFT developments based on the $SU(2)_{\mathcal{K}}$ current algebra for the explanation of TPS evolution in the overscreened Kondo regime; the exploration of similar physics for the $SU(N)_{\mathcal{K}}$ appears a valid avenue for further research. Yet another very interesting topic for future investigation might be to unveil the TPS evolution at the charge-Kondo regime.
%

\begin{thebibliography}{57}%
\makeatletter
\providecommand \@ifxundefined [1]{%
 \@ifx{#1\undefined}
}%
\providecommand \@ifnum [1]{%
 \ifnum #1\expandafter \@firstoftwo
 \else \expandafter \@secondoftwo
 \fi
}%
\providecommand \@ifx [1]{%
 \ifx #1\expandafter \@firstoftwo
 \else \expandafter \@secondoftwo
 \fi
}%
\providecommand \natexlab [1]{#1}%
\providecommand \enquote  [1]{``#1''}%
\providecommand \bibnamefont  [1]{#1}%
\providecommand \bibfnamefont [1]{#1}%
\providecommand \citenamefont [1]{#1}%
\providecommand \href@noop [0]{\@secondoftwo}%
\providecommand \href [0]{\begingroup \@sanitize@url \@href}%
\providecommand \@href[1]{\@@startlink{#1}\@@href}%
\providecommand \@@href[1]{\endgroup#1\@@endlink}%
\providecommand \@sanitize@url [0]{\catcode `\\12\catcode `\$12\catcode
  `\&12\catcode `\#12\catcode `\^12\catcode `\_12\catcode `\%12\relax}%
\providecommand \@@startlink[1]{}%
\providecommand \@@endlink[0]{}%
\providecommand \url  [0]{\begingroup\@sanitize@url \@url }%
\providecommand \@url [1]{\endgroup\@href {#1}{\urlprefix }}%
\providecommand \urlprefix  [0]{URL }%
\providecommand \Eprint [0]{\href }%
\providecommand \doibase [0]{http://dx.doi.org/}%
\providecommand \selectlanguage [0]{\@gobble}%
\providecommand \bibinfo  [0]{\@secondoftwo}%
\providecommand \bibfield  [0]{\@secondoftwo}%
\providecommand \translation [1]{[#1]}%
\providecommand \BibitemOpen [0]{}%
\providecommand \bibitemStop [0]{}%
\providecommand \bibitemNoStop [0]{.\EOS\space}%
\providecommand \EOS [0]{\spacefactor3000\relax}%
\providecommand \BibitemShut  [1]{\csname bibitem#1\endcsname}%
\let\auto@bib@innerbib\@empty
\bibitem [{\citenamefont {Kondo}(1964)}]{Kondo}%
  \BibitemOpen
  \bibfield  {author} {\bibinfo {author} {\bibfnamefont {J.}~\bibnamefont
  {Kondo}},\ }\href {http://dx.doi.org/10.1143/PTP.32.37} {\bibfield  {journal}
  {\bibinfo  {journal} {Prog. Theor. Phys.}\ }\textbf {\bibinfo {volume}
  {32}},\ \bibinfo {pages} {37} (\bibinfo {year} {1964})}\BibitemShut {NoStop}%
\bibitem [{\citenamefont {Anderson}(1970)}]{Anderson}%
  \BibitemOpen
  \bibfield  {author} {\bibinfo {author} {\bibfnamefont {P.~W.}\ \bibnamefont
  {Anderson}},\ }\href {http://stacks.iop.org/0022-3719/3/i=12/a=008}
  {\bibfield  {journal} {\bibinfo  {journal} {J. Phys. C}\ }\textbf {\bibinfo
  {volume} {3}},\ \bibinfo {pages} {2436} (\bibinfo {year} {1970})}\BibitemShut
  {NoStop}%
\bibitem [{\citenamefont {Yuval}\ and\ \citenamefont
  {Anderson}(1970)}]{haman1}%
  \BibitemOpen
  \bibfield  {author} {\bibinfo {author} {\bibfnamefont {G.}~\bibnamefont
  {Yuval}}\ and\ \bibinfo {author} {\bibfnamefont {P.~W.}\ \bibnamefont
  {Anderson}},\ }\href {\doibase 10.1103/PhysRevB.1.1522} {\bibfield  {journal}
  {\bibinfo  {journal} {Phys. Rev. B}\ }\textbf {\bibinfo {volume} {1}},\
  \bibinfo {pages} {1522} (\bibinfo {year} {1970})}\BibitemShut {NoStop}%
\bibitem [{\citenamefont {Anderson}\ \emph {et~al.}(1970)\citenamefont
  {Anderson}, \citenamefont {Yuval},\ and\ \citenamefont
  {Hamann}}]{Anderson_Yuval_Hamann}%
  \BibitemOpen
  \bibfield  {author} {\bibinfo {author} {\bibfnamefont {P.~W.}\ \bibnamefont
  {Anderson}}, \bibinfo {author} {\bibfnamefont {G.}~\bibnamefont {Yuval}}, \
  and\ \bibinfo {author} {\bibfnamefont {D.~R.}\ \bibnamefont {Hamann}},\
  }\href {https://journals.aps.org/prb/abstract/10.1103/PhysRevB.1.4464}
  {\bibfield  {journal} {\bibinfo  {journal} {Phys. Rev. B}\ }\textbf {\bibinfo
  {volume} {1}},\ \bibinfo {pages} {4464} (\bibinfo {year} {1970})}\BibitemShut
  {NoStop}%
\bibitem [{\citenamefont {Wilson}(1975)}]{Wilson}%
  \BibitemOpen
  \bibfield  {author} {\bibinfo {author} {\bibfnamefont {K.~G.}\ \bibnamefont
  {Wilson}},\ }\href {\doibase 10.1103/RevModPhys.47.773} {\bibfield  {journal}
  {\bibinfo  {journal} {Rev. Mod. Phys.}\ }\textbf {\bibinfo {volume} {47}},\
  \bibinfo {pages} {773} (\bibinfo {year} {1975})}\BibitemShut {NoStop}%
\bibitem [{\citenamefont {Tsvelik}\ and\ \citenamefont
  {Wiegmann}(1983)}]{jvm1}%
  \BibitemOpen
  \bibfield  {author} {\bibinfo {author} {\bibfnamefont {A.~M.}\ \bibnamefont
  {Tsvelik}}\ and\ \bibinfo {author} {\bibfnamefont {P.~B.}\ \bibnamefont
  {Wiegmann}},\ }\href {http://dx.doi.org/10.1080/00018738300101581} {\bibfield
   {journal} {\bibinfo  {journal} {Adv. Phys.}\ }\textbf {\bibinfo {volume}
  {32}},\ \bibinfo {pages} {453} (\bibinfo {year} {1983})}\BibitemShut
  {NoStop}%
\bibitem [{\citenamefont {Andrei}\ \emph {et~al.}(1983)\citenamefont {Andrei},
  \citenamefont {Furuya},\ and\ \citenamefont
  {Lowenstein}}]{Andrei_RevModPhys_1983}%
  \BibitemOpen
  \bibfield  {author} {\bibinfo {author} {\bibfnamefont {N.}~\bibnamefont
  {Andrei}}, \bibinfo {author} {\bibfnamefont {K.}~\bibnamefont {Furuya}}, \
  and\ \bibinfo {author} {\bibfnamefont {J.~H.}\ \bibnamefont {Lowenstein}},\
  }\href {\doibase 10.1103/RevModPhys.55.331} {\bibfield  {journal} {\bibinfo
  {journal} {Rev. Mod. Phys.}\ }\textbf {\bibinfo {volume} {55}},\ \bibinfo
  {pages} {331} (\bibinfo {year} {1983})}\BibitemShut {NoStop}%
\bibitem [{\citenamefont {Affleck}(1990)}]{AFFLECK_NPB_1990}%
  \BibitemOpen
  \bibfield  {author} {\bibinfo {author} {\bibfnamefont {I.}~\bibnamefont
  {Affleck}},\ }\href {\doibase https://doi.org/10.1016/0550-3213(90)90440-O}
  {\bibfield  {journal} {\bibinfo  {journal} {Nucl. Phys. B}\ }\textbf
  {\bibinfo {volume} {336}},\ \bibinfo {pages} {517 } (\bibinfo {year}
  {1990})}\BibitemShut {NoStop}%
\bibitem [{\citenamefont {Affleck}\ and\ \citenamefont
  {Ludwig}(1993)}]{Affleck_Lud_PRB(48)_1993}%
  \BibitemOpen
  \bibfield  {author} {\bibinfo {author} {\bibfnamefont {I.}~\bibnamefont
  {Affleck}}\ and\ \bibinfo {author} {\bibfnamefont {A.~W.~W.}\ \bibnamefont
  {Ludwig}},\ }\href
  {https://journals.aps.org/prb/abstract/10.1103/PhysRevB.48.7297} {\bibfield
  {journal} {\bibinfo  {journal} {Phys. Rev. B}\ }\textbf {\bibinfo {volume}
  {48}},\ \bibinfo {pages} {7297} (\bibinfo {year} {1993})}\BibitemShut
  {NoStop}%
\bibitem [{\citenamefont {Nozi{\'e}res}(1974)}]{Nozieres}%
  \BibitemOpen
  \bibfield  {author} {\bibinfo {author} {\bibfnamefont {P.}~\bibnamefont
  {Nozi{\'e}res}},\ }\href {https://doi.org/10.1007/BF00654541} {\bibfield
  {journal} {\bibinfo  {journal} {J. Low Temp. Phys.}\ }\textbf {\bibinfo
  {volume} {17}} (\bibinfo {year} {1974})}\BibitemShut {NoStop}%
\bibitem [{\citenamefont {Nozieres}\ and\ \citenamefont
  {Blandin}(1980)}]{Nozieres_Blandin_JPhys_1980}%
  \BibitemOpen
  \bibfield  {author} {\bibinfo {author} {\bibfnamefont {P.}~\bibnamefont
  {Nozieres}}\ and\ \bibinfo {author} {\bibfnamefont {A.}~\bibnamefont
  {Blandin}},\ }\href {https://doi.org/10.1051/jphys:01980004103019300}
  {\bibfield  {journal} {\bibinfo  {journal} {J. Phys}\ }\textbf {\bibinfo
  {volume} {41}},\ \bibinfo {pages} {193} (\bibinfo {year} {1980})}\BibitemShut
  {NoStop}%
\bibitem [{\citenamefont {Sacramento}\ and\ \citenamefont
  {Schlottmann}(1991)}]{Sacramento_CM(48)_1991}%
  \BibitemOpen
  \bibfield  {author} {\bibinfo {author} {\bibfnamefont {P.~D.}\ \bibnamefont
  {Sacramento}}\ and\ \bibinfo {author} {\bibfnamefont {P.}~\bibnamefont
  {Schlottmann}},\ }\href {http://stacks.iop.org/0953-8984/3/i=48/a=010}
  {\bibfield  {journal} {\bibinfo  {journal} {J. Phys.: Condens. Matter}\
  }\textbf {\bibinfo {volume} {3}},\ \bibinfo {pages} {9687} (\bibinfo {year}
  {1991})}\BibitemShut {NoStop}%
\bibitem [{\citenamefont {Cox}\ and\ \citenamefont
  {Zawadowski}(1998)}]{Cox_Adv_Phys(47)_1998}%
  \BibitemOpen
  \bibfield  {author} {\bibinfo {author} {\bibfnamefont {D.~L.}\ \bibnamefont
  {Cox}}\ and\ \bibinfo {author} {\bibfnamefont {A.}~\bibnamefont
  {Zawadowski}},\ }\href {http://dx.doi.org/10.1080/000187398243500} {\bibfield
   {journal} {\bibinfo  {journal} {Adv. Phys.}\ }\textbf {\bibinfo {volume}
  {47}},\ \bibinfo {pages} {599} (\bibinfo {year} {1998})}\BibitemShut
  {NoStop}%
\bibitem [{\citenamefont {Goldhaber-Gordon}\ \emph {et~al.}(1998)\citenamefont
  {Goldhaber-Gordon}, \citenamefont {Shtrikman}, \citenamefont {Mahalu},
  \citenamefont {Abusch-Magder}, \citenamefont {Meirav},\ and\ \citenamefont
  {Kastner}}]{Goldhaber_nat(391)_1998}%
  \BibitemOpen
  \bibfield  {author} {\bibinfo {author} {\bibfnamefont {D.}~\bibnamefont
  {Goldhaber-Gordon}}, \bibinfo {author} {\bibfnamefont {H.}~\bibnamefont
  {Shtrikman}}, \bibinfo {author} {\bibfnamefont {D.}~\bibnamefont {Mahalu}},
  \bibinfo {author} {\bibfnamefont {D.}~\bibnamefont {Abusch-Magder}}, \bibinfo
  {author} {\bibfnamefont {U.}~\bibnamefont {Meirav}}, \ and\ \bibinfo {author}
  {\bibfnamefont {M.~A.}\ \bibnamefont {Kastner}},\ }\href {\doibase
  10.1038/34373} {\bibfield  {journal} {\bibinfo  {journal} {Nature}\ }\textbf
  {\bibinfo {volume} {391}},\ \bibinfo {pages} {156} (\bibinfo {year}
  {1998})}\BibitemShut {NoStop}%
\bibitem [{\citenamefont {Oreg}\ and\ \citenamefont
  {Goldhaber-Gordon}(2003)}]{yuval_david}%
  \BibitemOpen
  \bibfield  {author} {\bibinfo {author} {\bibfnamefont {Y.}~\bibnamefont
  {Oreg}}\ and\ \bibinfo {author} {\bibfnamefont {D.}~\bibnamefont
  {Goldhaber-Gordon}},\ }\href {\doibase 10.1103/PhysRevLett.90.136602}
  {\bibfield  {journal} {\bibinfo  {journal} {Phys. Rev. Lett.}\ }\textbf
  {\bibinfo {volume} {90}},\ \bibinfo {pages} {136602} (\bibinfo {year}
  {2003})}\BibitemShut {NoStop}%
\bibitem [{\citenamefont {{Pustilnik}}\ \emph {et~al.}(2004)\citenamefont
  {{Pustilnik}}, \citenamefont {{Borda}}, \citenamefont {{Glazman}},\ and\
  \citenamefont {{von Delft}}}]{delft2}%
  \BibitemOpen
  \bibfield  {author} {\bibinfo {author} {\bibfnamefont {M.}~\bibnamefont
  {{Pustilnik}}}, \bibinfo {author} {\bibfnamefont {L.}~\bibnamefont
  {{Borda}}}, \bibinfo {author} {\bibfnamefont {L.~I.}\ \bibnamefont
  {{Glazman}}}, \ and\ \bibinfo {author} {\bibfnamefont {J.}~\bibnamefont {{von
  Delft}}},\ }\href {\doibase 10.1103/PhysRevB.69.115316} {\bibfield  {journal}
  {\bibinfo  {journal} {Phys. Rev. B}\ }\textbf {\bibinfo {volume} {69}},\
  \bibinfo {eid} {115316} (\bibinfo {year} {2004})}\BibitemShut {NoStop}%
\bibitem [{\citenamefont {Potok}\ \emph {et~al.}(2007)\citenamefont {Potok},
  \citenamefont {Rau}, \citenamefont {Shtrikman}, \citenamefont {Oreg},\ and\
  \citenamefont {Goldhaber-Gordon}}]{Potok_NAT(446)_2007}%
  \BibitemOpen
  \bibfield  {author} {\bibinfo {author} {\bibfnamefont {R.~M.}\ \bibnamefont
  {Potok}}, \bibinfo {author} {\bibfnamefont {I.~G.}\ \bibnamefont {Rau}},
  \bibinfo {author} {\bibfnamefont {H.}~\bibnamefont {Shtrikman}}, \bibinfo
  {author} {\bibfnamefont {Y.}~\bibnamefont {Oreg}}, \ and\ \bibinfo {author}
  {\bibfnamefont {D.}~\bibnamefont {Goldhaber-Gordon}},\ }\href
  {http://dx.doi.org/10.1038/nature05556} {\bibfield  {journal} {\bibinfo
  {journal} {Nature}\ }\textbf {\bibinfo {volume} {446}},\ \bibinfo {pages}
  {167} (\bibinfo {year} {2007})}\BibitemShut {NoStop}%
\bibitem [{\citenamefont {Hewson}(1993)}]{Hewson}%
  \BibitemOpen
  \bibfield  {author} {\bibinfo {author} {\bibfnamefont {A.}~\bibnamefont
  {Hewson}},\ }\href@noop {} {\emph {\bibinfo {title} {The Kondo Problem to
  Heavy Fermions}}}\ (\bibinfo  {publisher} {Cambridge University Press,
  Cambridge, England},\ \bibinfo {year} {1993})\BibitemShut {NoStop}%
\bibitem [{\citenamefont {Carmi}\ \emph {et~al.}(2012)\citenamefont {Carmi},
  \citenamefont {Oreg}, \citenamefont {Berkooz},\ and\ \citenamefont
  {Goldhaber-Gordon}}]{carmi1}%
  \BibitemOpen
  \bibfield  {author} {\bibinfo {author} {\bibfnamefont {A.}~\bibnamefont
  {Carmi}}, \bibinfo {author} {\bibfnamefont {Y.}~\bibnamefont {Oreg}},
  \bibinfo {author} {\bibfnamefont {M.}~\bibnamefont {Berkooz}}, \ and\
  \bibinfo {author} {\bibfnamefont {D.}~\bibnamefont {Goldhaber-Gordon}},\
  }\href {\doibase 10.1103/PhysRevB.86.115129} {\bibfield  {journal} {\bibinfo
  {journal} {Phys. Rev. B}\ }\textbf {\bibinfo {volume} {86}},\ \bibinfo
  {pages} {115129} (\bibinfo {year} {2012})}\BibitemShut {NoStop}%
\bibitem [{\citenamefont {Scheibner}\ \emph {et~al.}(2005)\citenamefont
  {Scheibner}, \citenamefont {Buhmann}, \citenamefont {Reuter}, \citenamefont
  {Kiselev},\ and\ \citenamefont {Molenkamp}}]{kiselev}%
  \BibitemOpen
  \bibfield  {author} {\bibinfo {author} {\bibfnamefont {R.}~\bibnamefont
  {Scheibner}}, \bibinfo {author} {\bibfnamefont {H.}~\bibnamefont {Buhmann}},
  \bibinfo {author} {\bibfnamefont {D.}~\bibnamefont {Reuter}}, \bibinfo
  {author} {\bibfnamefont {M.~N.}\ \bibnamefont {Kiselev}}, \ and\ \bibinfo
  {author} {\bibfnamefont {L.~W.}\ \bibnamefont {Molenkamp}},\ }\href {\doibase
  10.1103/PhysRevLett.95.176602} {\bibfield  {journal} {\bibinfo  {journal}
  {Phys. Rev. Lett.}\ }\textbf {\bibinfo {volume} {95}},\ \bibinfo {pages}
  {176602} (\bibinfo {year} {2005})}\BibitemShut {NoStop}%
\bibitem [{\citenamefont {Dutta}\ \emph {et~al.}(2019)\citenamefont {Dutta},
  \citenamefont {Majidi}, \citenamefont {Garcia~Corral}, \citenamefont
  {Erdman}, \citenamefont {Florens}, \citenamefont {Costi}, \citenamefont
  {Courtois},\ and\ \citenamefont {Winkelmann}}]{paulo}%
  \BibitemOpen
  \bibfield  {author} {\bibinfo {author} {\bibfnamefont {B.}~\bibnamefont
  {Dutta}}, \bibinfo {author} {\bibfnamefont {D.}~\bibnamefont {Majidi}},
  \bibinfo {author} {\bibfnamefont {A.}~\bibnamefont {Garcia~Corral}}, \bibinfo
  {author} {\bibfnamefont {P.~A.}\ \bibnamefont {Erdman}}, \bibinfo {author}
  {\bibfnamefont {S.}~\bibnamefont {Florens}}, \bibinfo {author} {\bibfnamefont
  {T.~A.}\ \bibnamefont {Costi}}, \bibinfo {author} {\bibfnamefont
  {H.}~\bibnamefont {Courtois}}, \ and\ \bibinfo {author} {\bibfnamefont
  {C.~B.}\ \bibnamefont {Winkelmann}},\ }\href {\doibase
  10.1021/acs.nanolett.8b04398} {\bibfield  {journal} {\bibinfo  {journal}
  {Nano Lett.}\ }\textbf {\bibinfo {volume} {19}},\ \bibinfo {pages} {506}
  (\bibinfo {year} {2019})}\BibitemShut {NoStop}%
\bibitem [{\citenamefont {Svilans}\ \emph {et~al.}(2018)\citenamefont
  {Svilans}, \citenamefont {Josefsson}, \citenamefont {Burke}, \citenamefont
  {Fahlvik}, \citenamefont {Thelander}, \citenamefont {Linke},\ and\
  \citenamefont {Leijnse}}]{heiner}%
  \BibitemOpen
  \bibfield  {author} {\bibinfo {author} {\bibfnamefont {A.}~\bibnamefont
  {Svilans}}, \bibinfo {author} {\bibfnamefont {M.}~\bibnamefont {Josefsson}},
  \bibinfo {author} {\bibfnamefont {A.~M.}\ \bibnamefont {Burke}}, \bibinfo
  {author} {\bibfnamefont {S.}~\bibnamefont {Fahlvik}}, \bibinfo {author}
  {\bibfnamefont {C.}~\bibnamefont {Thelander}}, \bibinfo {author}
  {\bibfnamefont {H.}~\bibnamefont {Linke}}, \ and\ \bibinfo {author}
  {\bibfnamefont {M.}~\bibnamefont {Leijnse}},\ }\href {\doibase
  10.1103/PhysRevLett.121.206801} {\bibfield  {journal} {\bibinfo  {journal}
  {Phys. Rev. Lett.}\ }\textbf {\bibinfo {volume} {121}},\ \bibinfo {pages}
  {206801} (\bibinfo {year} {2018})}\BibitemShut {NoStop}%
\bibitem [{\citenamefont {V.~Borzenets}\ \emph {et~al.}(2020)\citenamefont
  {V.~Borzenets}, \citenamefont {Shim}, \citenamefont {Chen}, \citenamefont
  {Ludwig}, \citenamefont {Wieck}, \citenamefont {Tarucha}, \citenamefont
  {Sim},\ and\ \citenamefont {Yamamoto}}]{new}%
  \BibitemOpen
  \bibfield  {author} {\bibinfo {author} {\bibfnamefont {I.}~\bibnamefont
  {V.~Borzenets}}, \bibinfo {author} {\bibfnamefont {J.}~\bibnamefont {Shim}},
  \bibinfo {author} {\bibfnamefont {J.~C.~H.}\ \bibnamefont {Chen}}, \bibinfo
  {author} {\bibfnamefont {A.}~\bibnamefont {Ludwig}}, \bibinfo {author}
  {\bibfnamefont {A.~D.}\ \bibnamefont {Wieck}}, \bibinfo {author}
  {\bibfnamefont {S.}~\bibnamefont {Tarucha}}, \bibinfo {author} {\bibfnamefont
  {H.-S.}\ \bibnamefont {Sim}}, \ and\ \bibinfo {author} {\bibfnamefont
  {M.}~\bibnamefont {Yamamoto}},\ }\href {\doibase 10.1038/s41586-020-2058-6}
  {\bibfield  {journal} {\bibinfo  {journal} {Nature}\ }\textbf {\bibinfo
  {volume} {579}},\ \bibinfo {pages} {210} (\bibinfo {year}
  {2020})}\BibitemShut {NoStop}%
\bibitem [{\citenamefont {Jarillo-Herrero}\ \emph {et~al.}(2005)\citenamefont
  {Jarillo-Herrero}, \citenamefont {Kong}, \citenamefont {van~der Zant},
  \citenamefont {Dekker}, \citenamefont {Kouwenhoven},\ and\ \citenamefont
  {Franceschi}}]{jh}%
  \BibitemOpen
  \bibfield  {author} {\bibinfo {author} {\bibfnamefont {P.}~\bibnamefont
  {Jarillo-Herrero}}, \bibinfo {author} {\bibfnamefont {J.}~\bibnamefont
  {Kong}}, \bibinfo {author} {\bibfnamefont {H.~S.}\ \bibnamefont {van~der
  Zant}}, \bibinfo {author} {\bibfnamefont {C.}~\bibnamefont {Dekker}},
  \bibinfo {author} {\bibfnamefont {L.~P.}\ \bibnamefont {Kouwenhoven}}, \ and\
  \bibinfo {author} {\bibfnamefont {S.~D.}\ \bibnamefont {Franceschi}},\ }\href
  {https://doi.org/10.1038/nature03422} {\bibfield  {journal} {\bibinfo
  {journal} {Nature}\ }\textbf {\bibinfo {volume} {434}},\ \bibinfo {pages}
  {484} (\bibinfo {year} {2005})}\BibitemShut {NoStop}%
\bibitem [{\citenamefont {Makarovski}\ \emph
  {et~al.}(2007{\natexlab{a}})\citenamefont {Makarovski}, \citenamefont {Liu},\
  and\ \citenamefont {Finkelstein}}]{sasa0}%
  \BibitemOpen
  \bibfield  {author} {\bibinfo {author} {\bibfnamefont {A.}~\bibnamefont
  {Makarovski}}, \bibinfo {author} {\bibfnamefont {J.}~\bibnamefont {Liu}}, \
  and\ \bibinfo {author} {\bibfnamefont {G.}~\bibnamefont {Finkelstein}},\
  }\href {\doibase 10.1103/PhysRevLett.99.066801} {\bibfield  {journal}
  {\bibinfo  {journal} {Phys. Rev. Lett.}\ }\textbf {\bibinfo {volume} {99}},\
  \bibinfo {pages} {066801} (\bibinfo {year} {2007}{\natexlab{a}})}\BibitemShut
  {NoStop}%
\bibitem [{\citenamefont {Makarovski}\ \emph
  {et~al.}(2007{\natexlab{b}})\citenamefont {Makarovski}, \citenamefont
  {Zhukov}, \citenamefont {Liu},\ and\ \citenamefont {Finkelstein}}]{sasa}%
  \BibitemOpen
  \bibfield  {author} {\bibinfo {author} {\bibfnamefont {A.}~\bibnamefont
  {Makarovski}}, \bibinfo {author} {\bibfnamefont {A.}~\bibnamefont {Zhukov}},
  \bibinfo {author} {\bibfnamefont {J.}~\bibnamefont {Liu}}, \ and\ \bibinfo
  {author} {\bibfnamefont {G.}~\bibnamefont {Finkelstein}},\ }\href {\doibase
  10.1103/PhysRevB.75.241407} {\bibfield  {journal} {\bibinfo  {journal} {Phys.
  Rev. B}\ }\textbf {\bibinfo {volume} {75}},\ \bibinfo {pages} {241407}
  (\bibinfo {year} {2007}{\natexlab{b}})}\BibitemShut {NoStop}%
\bibitem [{\citenamefont {Jezouin}\ \emph {et~al.}(2016)\citenamefont
  {Jezouin}, \citenamefont {Iftikhar}, \citenamefont {Anthore}, \citenamefont
  {Parmentier}, \citenamefont {Gennser}, \citenamefont {Cavanna}, \citenamefont
  {Ouerghi}, \citenamefont {Levkivskyi}, \citenamefont {Idrisov}, \citenamefont
  {Sukhorukov}, \citenamefont {Glazman},\ and\ \citenamefont {Pierre}}]{if2}%
  \BibitemOpen
  \bibfield  {author} {\bibinfo {author} {\bibfnamefont {S.}~\bibnamefont
  {Jezouin}}, \bibinfo {author} {\bibfnamefont {Z.}~\bibnamefont {Iftikhar}},
  \bibinfo {author} {\bibfnamefont {A.}~\bibnamefont {Anthore}}, \bibinfo
  {author} {\bibfnamefont {F.~D.}\ \bibnamefont {Parmentier}}, \bibinfo
  {author} {\bibfnamefont {U.}~\bibnamefont {Gennser}}, \bibinfo {author}
  {\bibfnamefont {A.}~\bibnamefont {Cavanna}}, \bibinfo {author} {\bibfnamefont
  {A.}~\bibnamefont {Ouerghi}}, \bibinfo {author} {\bibfnamefont {I.~P.}\
  \bibnamefont {Levkivskyi}}, \bibinfo {author} {\bibfnamefont
  {E.}~\bibnamefont {Idrisov}}, \bibinfo {author} {\bibfnamefont {E.~V.}\
  \bibnamefont {Sukhorukov}}, \bibinfo {author} {\bibfnamefont {L.~I.}\
  \bibnamefont {Glazman}}, \ and\ \bibinfo {author} {\bibfnamefont
  {F.}~\bibnamefont {Pierre}},\ }\href {http://dx.doi.org/10.1038/nature19072}
  {\bibfield  {journal} {\bibinfo  {journal} {Nature}\ }\textbf {\bibinfo
  {volume} {536}},\ \bibinfo {pages} {60} (\bibinfo {year} {2016})}\BibitemShut
  {NoStop}%
\bibitem [{\citenamefont {Hata}\ \emph {et~al.}(2018)\citenamefont {Hata},
  \citenamefont {Delagrange}, \citenamefont {Arakawa}, \citenamefont {Lee},
  \citenamefont {Deblock}, \citenamefont {Bouchiat}, \citenamefont
  {Kobayashi},\ and\ \citenamefont {Ferrier}}]{ll1}%
  \BibitemOpen
  \bibfield  {author} {\bibinfo {author} {\bibfnamefont {T.}~\bibnamefont
  {Hata}}, \bibinfo {author} {\bibfnamefont {R.}~\bibnamefont {Delagrange}},
  \bibinfo {author} {\bibfnamefont {T.}~\bibnamefont {Arakawa}}, \bibinfo
  {author} {\bibfnamefont {S.}~\bibnamefont {Lee}}, \bibinfo {author}
  {\bibfnamefont {R.}~\bibnamefont {Deblock}}, \bibinfo {author} {\bibfnamefont
  {H.}~\bibnamefont {Bouchiat}}, \bibinfo {author} {\bibfnamefont
  {K.}~\bibnamefont {Kobayashi}}, \ and\ \bibinfo {author} {\bibfnamefont
  {M.}~\bibnamefont {Ferrier}},\ }\href {\doibase
  10.1103/PhysRevLett.121.247703} {\bibfield  {journal} {\bibinfo  {journal}
  {Phys. Rev. Lett.}\ }\textbf {\bibinfo {volume} {121}},\ \bibinfo {pages}
  {247703} (\bibinfo {year} {2018})}\BibitemShut {NoStop}%
\bibitem [{\citenamefont {Keller}\ \emph {et~al.}(2014)\citenamefont {Keller},
  \citenamefont {Amasha}, \citenamefont {Weymann}, \citenamefont {Moca},
  \citenamefont {Rau}, \citenamefont {Katine}, \citenamefont {Shtrikman},
  \citenamefont {Zarand},\ and\ \citenamefont {Goldhaber-Gordon}}]{keller}%
  \BibitemOpen
  \bibfield  {author} {\bibinfo {author} {\bibfnamefont {A.~J.}\ \bibnamefont
  {Keller}}, \bibinfo {author} {\bibfnamefont {S.}~\bibnamefont {Amasha}},
  \bibinfo {author} {\bibfnamefont {I.}~\bibnamefont {Weymann}}, \bibinfo
  {author} {\bibfnamefont {C.~P.}\ \bibnamefont {Moca}}, \bibinfo {author}
  {\bibfnamefont {I.~G.}\ \bibnamefont {Rau}}, \bibinfo {author} {\bibfnamefont
  {J.~A.}\ \bibnamefont {Katine}}, \bibinfo {author} {\bibfnamefont
  {H.}~\bibnamefont {Shtrikman}}, \bibinfo {author} {\bibfnamefont
  {G.}~\bibnamefont {Zarand}}, \ and\ \bibinfo {author} {\bibfnamefont
  {D.}~\bibnamefont {Goldhaber-Gordon}},\ }\href
  {http://dx.doi.org/10.1038/nphys2844} {\bibfield  {journal} {\bibinfo
  {journal} {Nat. Phys.}\ }\textbf {\bibinfo {volume} {10}},\ \bibinfo {pages}
  {145} (\bibinfo {year} {2014})}\BibitemShut {NoStop}%
\bibitem [{\citenamefont {Le~Hur}\ \emph {et~al.}(2007)\citenamefont {Le~Hur},
  \citenamefont {Simon},\ and\ \citenamefont {Loss}}]{hur}%
  \BibitemOpen
  \bibfield  {author} {\bibinfo {author} {\bibfnamefont {K.}~\bibnamefont
  {Le~Hur}}, \bibinfo {author} {\bibfnamefont {P.}~\bibnamefont {Simon}}, \
  and\ \bibinfo {author} {\bibfnamefont {D.}~\bibnamefont {Loss}},\ }\href
  {\doibase 10.1103/PhysRevB.75.035332} {\bibfield  {journal} {\bibinfo
  {journal} {Phys. Rev. B}\ }\textbf {\bibinfo {volume} {75}},\ \bibinfo
  {pages} {035332} (\bibinfo {year} {2007})}\BibitemShut {NoStop}%
\bibitem [{\citenamefont {Choi}\ \emph {et~al.}(2005)\citenamefont {Choi},
  \citenamefont {L\'opez},\ and\ \citenamefont {Aguado}}]{su4_21}%
  \BibitemOpen
  \bibfield  {author} {\bibinfo {author} {\bibfnamefont {M.-S.}\ \bibnamefont
  {Choi}}, \bibinfo {author} {\bibfnamefont {R.}~\bibnamefont {L\'opez}}, \
  and\ \bibinfo {author} {\bibfnamefont {R.}~\bibnamefont {Aguado}},\ }\href
  {\doibase 10.1103/PhysRevLett.95.067204} {\bibfield  {journal} {\bibinfo
  {journal} {Phys. Rev. Lett.}\ }\textbf {\bibinfo {volume} {95}},\ \bibinfo
  {pages} {067204} (\bibinfo {year} {2005})}\BibitemShut {NoStop}%
\bibitem [{\citenamefont {Lim}\ \emph {et~al.}(2006)\citenamefont {Lim},
  \citenamefont {Choi}, \citenamefont {Choi}, \citenamefont {L\'opez},\ and\
  \citenamefont {Aguado}}]{su4_23}%
  \BibitemOpen
  \bibfield  {author} {\bibinfo {author} {\bibfnamefont {J.~S.}\ \bibnamefont
  {Lim}}, \bibinfo {author} {\bibfnamefont {M.-S.}\ \bibnamefont {Choi}},
  \bibinfo {author} {\bibfnamefont {M.~Y.}\ \bibnamefont {Choi}}, \bibinfo
  {author} {\bibfnamefont {R.}~\bibnamefont {L\'opez}}, \ and\ \bibinfo
  {author} {\bibfnamefont {R.}~\bibnamefont {Aguado}},\ }\href {\doibase
  10.1103/PhysRevB.74.205119} {\bibfield  {journal} {\bibinfo  {journal} {Phys.
  Rev. B}\ }\textbf {\bibinfo {volume} {74}},\ \bibinfo {pages} {205119}
  (\bibinfo {year} {2006})}\BibitemShut {NoStop}%
\bibitem [{\citenamefont {Lim}\ \emph {et~al.}(2014)\citenamefont {Lim},
  \citenamefont {L{\'{o}}pez},\ and\ \citenamefont {S{\'{a}}nchez}}]{Lim}%
  \BibitemOpen
  \bibfield  {author} {\bibinfo {author} {\bibfnamefont {J.~S.}\ \bibnamefont
  {Lim}}, \bibinfo {author} {\bibfnamefont {R.}~\bibnamefont {L{\'{o}}pez}}, \
  and\ \bibinfo {author} {\bibfnamefont {D.}~\bibnamefont {S{\'{a}}nchez}},\
  }\href {\doibase 10.1088/1367-2630/16/1/015003} {\bibfield  {journal}
  {\bibinfo  {journal} {New J. Phys.}\ }\textbf {\bibinfo {volume} {16}},\
  \bibinfo {pages} {015003} (\bibinfo {year} {2014})}\BibitemShut {NoStop}%
\bibitem [{\citenamefont {Kleeorin}\ and\ \citenamefont {Meir}(2017)}]{su4_11}%
  \BibitemOpen
  \bibfield  {author} {\bibinfo {author} {\bibfnamefont {Y.}~\bibnamefont
  {Kleeorin}}\ and\ \bibinfo {author} {\bibfnamefont {Y.}~\bibnamefont
  {Meir}},\ }\href {\doibase 10.1103/PhysRevB.96.045118} {\bibfield  {journal}
  {\bibinfo  {journal} {Phys. Rev. B}\ }\textbf {\bibinfo {volume} {96}},\
  \bibinfo {pages} {045118} (\bibinfo {year} {2017})}\BibitemShut {NoStop}%
\bibitem [{\citenamefont {Karki}\ and\ \citenamefont {Kiselev}(2017)}]{dee1}%
  \BibitemOpen
  \bibfield  {author} {\bibinfo {author} {\bibfnamefont {D.~B.}\ \bibnamefont
  {Karki}}\ and\ \bibinfo {author} {\bibfnamefont {M.~N.}\ \bibnamefont
  {Kiselev}},\ }\href {\doibase 10.1103/PhysRevB.96.121403} {\bibfield
  {journal} {\bibinfo  {journal} {Phys. Rev. B}\ }\textbf {\bibinfo {volume}
  {96}},\ \bibinfo {pages} {121403(R)} (\bibinfo {year} {2017})}\BibitemShut
  {NoStop}%
\bibitem [{\citenamefont {Karki}\ and\ \citenamefont {Kiselev}(2019)}]{last1}%
  \BibitemOpen
  \bibfield  {author} {\bibinfo {author} {\bibfnamefont {D.~B.}\ \bibnamefont
  {Karki}}\ and\ \bibinfo {author} {\bibfnamefont {M.~N.}\ \bibnamefont
  {Kiselev}},\ }\href {\doibase 10.1103/PhysRevB.100.125426} {\bibfield
  {journal} {\bibinfo  {journal} {Phys. Rev. B}\ }\textbf {\bibinfo {volume}
  {100}},\ \bibinfo {pages} {125426} (\bibinfo {year} {2019})}\BibitemShut
  {NoStop}%
\bibitem [{\citenamefont {{Karki}}\ and\ \citenamefont
  {{Kiselev}}(2019)}]{den}%
  \BibitemOpen
  \bibfield  {author} {\bibinfo {author} {\bibfnamefont {D.~B.}\ \bibnamefont
  {{Karki}}}\ and\ \bibinfo {author} {\bibfnamefont {M.~N.}\ \bibnamefont
  {{Kiselev}}},\ }\href {\doibase 10.1103/PhysRevB.100.195425} {\bibfield
  {journal} {\bibinfo  {journal} {Phys. Rev. B}\ }\textbf {\bibinfo {volume}
  {100}},\ \bibinfo {eid} {195425} (\bibinfo {year} {2019})}\BibitemShut
  {NoStop}%
\bibitem [{\citenamefont {Iftikhar}\ \emph {et~al.}(2015)\citenamefont
  {Iftikhar}, \citenamefont {Jezouin}, \citenamefont {Anthore}, \citenamefont
  {Gennser}, \citenamefont {Parmentier}, \citenamefont {Cavanna},\ and\
  \citenamefont {Pierre}}]{Pierre_NAT(526)_2015}%
  \BibitemOpen
  \bibfield  {author} {\bibinfo {author} {\bibfnamefont {Z.}~\bibnamefont
  {Iftikhar}}, \bibinfo {author} {\bibfnamefont {S.}~\bibnamefont {Jezouin}},
  \bibinfo {author} {\bibfnamefont {A.}~\bibnamefont {Anthore}}, \bibinfo
  {author} {\bibfnamefont {U.}~\bibnamefont {Gennser}}, \bibinfo {author}
  {\bibfnamefont {F.~D.}\ \bibnamefont {Parmentier}}, \bibinfo {author}
  {\bibfnamefont {A.}~\bibnamefont {Cavanna}}, \ and\ \bibinfo {author}
  {\bibfnamefont {F.}~\bibnamefont {Pierre}},\ }\href
  {http://dx.doi.org/10.1038/nature15384} {\bibfield  {journal} {\bibinfo
  {journal} {Nature}\ }\textbf {\bibinfo {volume} {526}},\ \bibinfo {pages}
  {233} (\bibinfo {year} {2015})}\BibitemShut {NoStop}%
\bibitem [{\citenamefont {{Iftikhar}}\ \emph {et~al.}(2018)\citenamefont
  {{Iftikhar}}, \citenamefont {{Anthore}}, \citenamefont {{Mitchell}},
  \citenamefont {{Parmentier}}, \citenamefont {{Gennser}}, \citenamefont
  {{Ouerghi}}, \citenamefont {{Cavanna}}, \citenamefont {{Mora}}, \citenamefont
  {{Simon}},\ and\ \citenamefont {{Pierre}}}]{iff}%
  \BibitemOpen
  \bibfield  {author} {\bibinfo {author} {\bibfnamefont {Z.}~\bibnamefont
  {{Iftikhar}}}, \bibinfo {author} {\bibfnamefont {A.}~\bibnamefont
  {{Anthore}}}, \bibinfo {author} {\bibfnamefont {A.~K.}\ \bibnamefont
  {{Mitchell}}}, \bibinfo {author} {\bibfnamefont {F.~D.}\ \bibnamefont
  {{Parmentier}}}, \bibinfo {author} {\bibfnamefont {U.}~\bibnamefont
  {{Gennser}}}, \bibinfo {author} {\bibfnamefont {A.}~\bibnamefont
  {{Ouerghi}}}, \bibinfo {author} {\bibfnamefont {A.}~\bibnamefont
  {{Cavanna}}}, \bibinfo {author} {\bibfnamefont {C.}~\bibnamefont {{Mora}}},
  \bibinfo {author} {\bibfnamefont {P.}~\bibnamefont {{Simon}}}, \ and\
  \bibinfo {author} {\bibfnamefont {F.}~\bibnamefont {{Pierre}}},\ }\href
  {\doibase 10.1126/science.aan5592} {\bibfield  {journal} {\bibinfo  {journal}
  {Science}\ }\textbf {\bibinfo {volume} {360}},\ \bibinfo {pages} {1315}
  (\bibinfo {year} {2018})}\BibitemShut {NoStop}%
\bibitem [{\citenamefont {Ji}\ \emph {et~al.}(2002)\citenamefont {Ji},
  \citenamefont {Heiblum},\ and\ \citenamefont {Shtrikman}}]{moti_hiblum}%
  \BibitemOpen
  \bibfield  {author} {\bibinfo {author} {\bibfnamefont {Y.}~\bibnamefont
  {Ji}}, \bibinfo {author} {\bibfnamefont {M.}~\bibnamefont {Heiblum}}, \ and\
  \bibinfo {author} {\bibfnamefont {H.}~\bibnamefont {Shtrikman}},\ }\href
  {\doibase 10.1103/PhysRevLett.88.076601} {\bibfield  {journal} {\bibinfo
  {journal} {Phys. Rev. Lett.}\ }\textbf {\bibinfo {volume} {88}},\ \bibinfo
  {pages} {076601} (\bibinfo {year} {2002})}\BibitemShut {NoStop}%
\bibitem [{\citenamefont {Ji}\ \emph {et~al.}(2000)\citenamefont {Ji},
  \citenamefont {Heiblum}, \citenamefont {Sprinzak}, \citenamefont {Mahalu},\
  and\ \citenamefont {Shtrikman}}]{moti_1}%
  \BibitemOpen
  \bibfield  {author} {\bibinfo {author} {\bibfnamefont {Y.}~\bibnamefont
  {Ji}}, \bibinfo {author} {\bibfnamefont {M.}~\bibnamefont {Heiblum}},
  \bibinfo {author} {\bibfnamefont {D.}~\bibnamefont {Sprinzak}}, \bibinfo
  {author} {\bibfnamefont {D.}~\bibnamefont {Mahalu}}, \ and\ \bibinfo {author}
  {\bibfnamefont {H.}~\bibnamefont {Shtrikman}},\ }\href {\doibase
  10.1126/science.290.5492.779} {\bibfield  {journal} {\bibinfo  {journal}
  {Science}\ }\textbf {\bibinfo {volume} {290}},\ \bibinfo {pages} {779}
  (\bibinfo {year} {2000})}\BibitemShut {NoStop}%
\bibitem [{\citenamefont {Zaffalon}\ \emph {et~al.}(2008)\citenamefont
  {Zaffalon}, \citenamefont {Bid}, \citenamefont {Heiblum}, \citenamefont
  {Mahalu},\ and\ \citenamefont {Umansky}}]{moti_2008}%
  \BibitemOpen
  \bibfield  {author} {\bibinfo {author} {\bibfnamefont {M.}~\bibnamefont
  {Zaffalon}}, \bibinfo {author} {\bibfnamefont {A.}~\bibnamefont {Bid}},
  \bibinfo {author} {\bibfnamefont {M.}~\bibnamefont {Heiblum}}, \bibinfo
  {author} {\bibfnamefont {D.}~\bibnamefont {Mahalu}}, \ and\ \bibinfo {author}
  {\bibfnamefont {V.}~\bibnamefont {Umansky}},\ }\href {\doibase
  10.1103/PhysRevLett.100.226601} {\bibfield  {journal} {\bibinfo  {journal}
  {Phys. Rev. Lett.}\ }\textbf {\bibinfo {volume} {100}},\ \bibinfo {pages}
  {226601} (\bibinfo {year} {2008})}\BibitemShut {NoStop}%
\bibitem [{\citenamefont {Takada}\ \emph {et~al.}(2014)\citenamefont {Takada},
  \citenamefont {B\"auerle}, \citenamefont {Yamamoto}, \citenamefont
  {Watanabe}, \citenamefont {Hermelin}, \citenamefont {Meunier}, \citenamefont
  {Alex}, \citenamefont {Weichselbaum}, \citenamefont {von Delft},
  \citenamefont {Ludwig}, \citenamefont {Wieck},\ and\ \citenamefont
  {Tarucha}}]{delft_andrea}%
  \BibitemOpen
  \bibfield  {author} {\bibinfo {author} {\bibfnamefont {S.}~\bibnamefont
  {Takada}}, \bibinfo {author} {\bibfnamefont {C.}~\bibnamefont {B\"auerle}},
  \bibinfo {author} {\bibfnamefont {M.}~\bibnamefont {Yamamoto}}, \bibinfo
  {author} {\bibfnamefont {K.}~\bibnamefont {Watanabe}}, \bibinfo {author}
  {\bibfnamefont {S.}~\bibnamefont {Hermelin}}, \bibinfo {author}
  {\bibfnamefont {T.}~\bibnamefont {Meunier}}, \bibinfo {author} {\bibfnamefont
  {A.}~\bibnamefont {Alex}}, \bibinfo {author} {\bibfnamefont {A.}~\bibnamefont
  {Weichselbaum}}, \bibinfo {author} {\bibfnamefont {J.}~\bibnamefont {von
  Delft}}, \bibinfo {author} {\bibfnamefont {A.}~\bibnamefont {Ludwig}},
  \bibinfo {author} {\bibfnamefont {A.~D.}\ \bibnamefont {Wieck}}, \ and\
  \bibinfo {author} {\bibfnamefont {S.}~\bibnamefont {Tarucha}},\ }\href
  {\doibase 10.1103/PhysRevLett.113.126601} {\bibfield  {journal} {\bibinfo
  {journal} {Phys. Rev. Lett.}\ }\textbf {\bibinfo {volume} {113}},\ \bibinfo
  {pages} {126601} (\bibinfo {year} {2014})}\BibitemShut {NoStop}%
\bibitem [{Note1()}]{Note1}%
  \BibitemOpen
  \bibinfo {note} {The presented Fermi-liquid description is limited to the
  single impurity Anderson model with symmetric bath density of
  states.}\BibitemShut {Stop}%
\bibitem [{\citenamefont {Affleck}(2005)}]{affjs}%
  \BibitemOpen
  \bibfield  {author} {\bibinfo {author} {\bibfnamefont {I.}~\bibnamefont
  {Affleck}},\ }\href {\doibase 10.1143/JPSJ.74.59} {\bibfield  {journal}
  {\bibinfo  {journal} {J. Phys. Soc. Jpn.}\ }\textbf
  {\bibinfo {volume} {74}},\ \bibinfo {pages} {59} (\bibinfo {year}
  {2005})}\BibitemShut {NoStop}%
\bibitem [{\citenamefont {Gerland}\ \emph {et~al.}(2000)\citenamefont
  {Gerland}, \citenamefont {von Delft}, \citenamefont {Costi},\ and\
  \citenamefont {Oreg}}]{oreg_delft}%
  \BibitemOpen
  \bibfield  {author} {\bibinfo {author} {\bibfnamefont {U.}~\bibnamefont
  {Gerland}}, \bibinfo {author} {\bibfnamefont {J.}~\bibnamefont {von Delft}},
  \bibinfo {author} {\bibfnamefont {T.~A.}\ \bibnamefont {Costi}}, \ and\
  \bibinfo {author} {\bibfnamefont {Y.}~\bibnamefont {Oreg}},\ }\href {\doibase
  10.1103/PhysRevLett.84.3710} {\bibfield  {journal} {\bibinfo  {journal}
  {Phys. Rev. Lett.}\ }\textbf {\bibinfo {volume} {84}},\ \bibinfo {pages}
  {3710} (\bibinfo {year} {2000})}\BibitemShut {NoStop}%
\bibitem [{\citenamefont {Hanl}\ \emph {et~al.}(2014)\citenamefont {Hanl},
  \citenamefont {Weichselbaum}, \citenamefont {von Delft},\ and\ \citenamefont
  {Kiselev}}]{HWDK_PRB_(89)_2014}%
  \BibitemOpen
  \bibfield  {author} {\bibinfo {author} {\bibfnamefont {M.}~\bibnamefont
  {Hanl}}, \bibinfo {author} {\bibfnamefont {A.}~\bibnamefont {Weichselbaum}},
  \bibinfo {author} {\bibfnamefont {J.}~\bibnamefont {von Delft}}, \ and\
  \bibinfo {author} {\bibfnamefont {M.}~\bibnamefont {Kiselev}},\ }\href
  {https://journals.aps.org/prb/abstract/10.1103/PhysRevB.89.195131} {\bibfield
   {journal} {\bibinfo  {journal} {Phys. Rev. B}\ }\textbf {\bibinfo {volume}
  {89}},\ \bibinfo {pages} {195131} (\bibinfo {year} {2014})}\BibitemShut
  {NoStop}%
\bibitem [{\citenamefont {Karki}\ \emph {et~al.}(2018)\citenamefont {Karki},
  \citenamefont {Mora}, \citenamefont {von Delft},\ and\ \citenamefont
  {Kiselev}}]{dee2}%
  \BibitemOpen
  \bibfield  {author} {\bibinfo {author} {\bibfnamefont {D.~B.}\ \bibnamefont
  {Karki}}, \bibinfo {author} {\bibfnamefont {C.}~\bibnamefont {Mora}},
  \bibinfo {author} {\bibfnamefont {J.}~\bibnamefont {von Delft}}, \ and\
  \bibinfo {author} {\bibfnamefont {M.~N.}\ \bibnamefont {Kiselev}},\ }\href
  {\doibase 10.1103/PhysRevB.97.195403} {\bibfield  {journal} {\bibinfo
  {journal} {Phys. Rev. B}\ }\textbf {\bibinfo {volume} {97}},\ \bibinfo
  {pages} {195403} (\bibinfo {year} {2018})}\BibitemShut {NoStop}%
\bibitem [{\citenamefont {Karki}\ and\ \citenamefont {Kiselev}(2018)}]{dee3}%
  \BibitemOpen
  \bibfield  {author} {\bibinfo {author} {\bibfnamefont {D.~B.}\ \bibnamefont
  {Karki}}\ and\ \bibinfo {author} {\bibfnamefont {M.~N.}\ \bibnamefont
  {Kiselev}},\ }\href {\doibase 10.1103/PhysRevB.98.165443} {\bibfield
  {journal} {\bibinfo  {journal} {Phys. Rev. B}\ }\textbf {\bibinfo {volume}
  {98}},\ \bibinfo {pages} {165443} (\bibinfo {year} {2018})}\BibitemShut
  {NoStop}%
\bibitem [{Note2()}]{Note2}%
  \BibitemOpen
  \bibinfo {note} {We note that the presented work describes the case of
  filling factor $m/N<1$ with $m=1, 2,\protect \cdots N-1$.}\BibitemShut
  {Stop}%
\bibitem [{\citenamefont {Anderson}(1961)}]{and1}%
  \BibitemOpen
  \bibfield  {author} {\bibinfo {author} {\bibfnamefont {P.~W.}\ \bibnamefont
  {Anderson}},\ }\href {\doibase 10.1103/PhysRev.124.41} {\bibfield  {journal}
  {\bibinfo  {journal} {Phys. Rev.}\ }\textbf {\bibinfo {volume} {124}},\
  \bibinfo {pages} {41} (\bibinfo {year} {1961})}\BibitemShut {NoStop}%
\bibitem [{\citenamefont {Pustilnik}\ and\ \citenamefont
  {Glazman}(2004)}]{GP_Review_2005}%
  \BibitemOpen
  \bibfield  {author} {\bibinfo {author} {\bibfnamefont {M.}~\bibnamefont
  {Pustilnik}}\ and\ \bibinfo {author} {\bibfnamefont {L.}~\bibnamefont
  {Glazman}},\ }\href {http://stacks.iop.org/0953-8984/16/i=16/a=R01}
  {\bibfield  {journal} {\bibinfo  {journal} {J. Phys.: Condens. Matter}\
  }\textbf {\bibinfo {volume} {16}},\ \bibinfo {pages} {R513} (\bibinfo {year}
  {2004})}\BibitemShut {NoStop}%
\bibitem [{\citenamefont {Mora}\ \emph {et~al.}(2009)\citenamefont {Mora},
  \citenamefont {Vitushinsky}, \citenamefont {Leyronas}, \citenamefont
  {Clerk},\ and\ \citenamefont {Le~Hur}}]{mora1}%
  \BibitemOpen
  \bibfield  {author} {\bibinfo {author} {\bibfnamefont {C.}~\bibnamefont
  {Mora}}, \bibinfo {author} {\bibfnamefont {P.}~\bibnamefont {Vitushinsky}},
  \bibinfo {author} {\bibfnamefont {X.}~\bibnamefont {Leyronas}}, \bibinfo
  {author} {\bibfnamefont {A.~A.}\ \bibnamefont {Clerk}}, \ and\ \bibinfo
  {author} {\bibfnamefont {K.}~\bibnamefont {Le~Hur}},\ }\href {\doibase
  10.1103/PhysRevB.80.155322} {\bibfield  {journal} {\bibinfo  {journal} {Phys.
  Rev. B}\ }\textbf {\bibinfo {volume} {80}},\ \bibinfo {pages} {155322}
  (\bibinfo {year} {2009})}\BibitemShut {NoStop}%
\bibitem [{\citenamefont {Mora}(2009)}]{mora2}%
  \BibitemOpen
  \bibfield  {author} {\bibinfo {author} {\bibfnamefont {C.}~\bibnamefont
  {Mora}},\ }\href
  {https://journals.aps.org/prb/abstract/10.1103/PhysRevB.80.125304} {\bibfield
   {journal} {\bibinfo  {journal} {Phys. Rev. B}\ }\textbf {\bibinfo {volume}
  {80}},\ \bibinfo {pages} {125304} (\bibinfo {year} {2009})}\BibitemShut
  {NoStop}%
\bibitem [{\citenamefont {Mora}\ \emph {et~al.}(2015)\citenamefont {Mora},
  \citenamefont {Moca}, \citenamefont {von Delft},\ and\ \citenamefont
  {Zar\'and}}]{jvm}%
  \BibitemOpen
  \bibfield  {author} {\bibinfo {author} {\bibfnamefont {C.}~\bibnamefont
  {Mora}}, \bibinfo {author} {\bibfnamefont {C.~P.}\ \bibnamefont {Moca}},
  \bibinfo {author} {\bibfnamefont {J.}~\bibnamefont {von Delft}}, \ and\
  \bibinfo {author} {\bibfnamefont {G.}~\bibnamefont {Zar\'and}},\ }\href
  {\doibase 10.1103/PhysRevB.92.075120} {\bibfield  {journal} {\bibinfo
  {journal} {Phys. Rev. B}\ }\textbf {\bibinfo {volume} {92}},\ \bibinfo
  {pages} {075120} (\bibinfo {year} {2015})}\BibitemShut {NoStop}%
\bibitem [{\citenamefont {Karki}\ and\ \citenamefont {Kiselev}(2020)}]{dbk}%
  \BibitemOpen
  \bibfield  {author} {\bibinfo {author} {\bibfnamefont {D.~B.}\ \bibnamefont
  {Karki}}\ and\ \bibinfo {author} {\bibfnamefont {M.~N.}\ \bibnamefont
  {Kiselev}},\ }\href {\doibase 10.1103/PhysRevB.102.241402} {\bibfield
  {journal} {\bibinfo  {journal} {Phys. Rev. B}\ }\textbf {\bibinfo {volume}
  {102}},\ \bibinfo {pages} {241402} (\bibinfo {year} {2020})}\BibitemShut
  {NoStop}%
\bibitem [{\citenamefont {B\'eri}\ and\ \citenamefont {Cooper}(2012)}]{beri}%
  \BibitemOpen
  \bibfield  {author} {\bibinfo {author} {\bibfnamefont {B.}~\bibnamefont
  {B\'eri}}\ and\ \bibinfo {author} {\bibfnamefont {N.~R.}\ \bibnamefont
  {Cooper}},\ }\href {\doibase 10.1103/PhysRevLett.109.156803} {\bibfield
  {journal} {\bibinfo  {journal} {Phys. Rev. Lett.}\ }\textbf {\bibinfo
  {volume} {109}},\ \bibinfo {pages} {156803} (\bibinfo {year}
  {2012})}\BibitemShut {NoStop}%
\end{thebibliography}
\end{document}